\documentstyle[12pt,epsfig]{article}

\newcommand{\beeq}{\begin{equation}}
\newcommand{\eneq}{\end{equation}}
\newcommand{\beeqar}{\begin{eqnarray}}
\newcommand{\eneqar}{\end{eqnarray}}
\begin{document}
\vskip 0.5in
\begin{center}
{\large \bf Asymptotically Free Non-Abelian Gauge Theories With Fermions
and Scalars As Alternatives to QCD}\\
\vskip 0.5in
N.D. Hari Dass\dag\\ 
 Institute of Mathematical Sciences, C.I.T Campus, Chennai 600 113, INDIA\\
 V. Soni\ddag\\
 National Physical Laboratory, New Delhi, INDIA
\end{center}
\vskip 0.5in
{\large \bf Abstract}\\
In this paper we construct non-Abelian gauge theories with fermions and
scalars that nevertheless possess asymptotic freedom.The scalars are
taken to be in a chiral multiplet transforming as $(2,2)$ under
$SU(2)_L\otimes SU(2)_R$ and transforming as singlets under the colour
$SU(3)$ group. We consider two distinct scenarios, one in which the
additional scalars are light and another in which they are heavier than
half the Z-boson mass. It is shown that asymptotic freedom is obtained
without requiring that all additional couplings keep fixed ratios with
each other. It is also shown that both scenarios can not be ruled out
by what are considered standard tests of QCD like R- parameter, g-2 for
muons or deep inelastic phenomena. The light mass scenario is however
ruled out by high precision Z-width data (and only by that one data).The
heavy mass scenario is still viable and is shown to naturally pass the
test of flavour changing neutral currents.It also is not ruled out by
precision electroweak oblique parameters.Many distinctive experimental
signatures of these models are also discussed.
\section{Introduction}
Quantum chromodynamics (QCD) is today believed to be the correct theory of strong interactions. The reasons for this belief are, on the one hand, the fact
that the global symmetries of QCD and that of the observed hadronic world are
the same, and on the other hand, the phenomenon of asymptotic freedom(AF)
\cite{free}\footnote { Seiler and Patrasciou \cite{seiler}have, however, presented a
dissenting perspective on Asymptotic Freedom and reliability of
perturbative
methods in non-abelian gauge theories }.
Indeed,
asymptotic freedom (AF) allows one to make experimental comparisons with the predictions of
QCD in the deep inelastic regime and the evidence in favour of QCD appears 
to be quite compelling. Nevertheless, it is extremely important to analyse the
experimental data within a broader theoretical framework that contains QCD as
a special case.

The need for a framework to analyse data that is larger than the theory one
is
trying to uphold is fairly obvious. Otherwise the data analysis is likely
to be biased by elements of the theory itself. Of course, no analysis of the
data is possible without some theoretical framework. But such a theoretical
framework should be as broad as possible. For example, analysis of
gravitational phenomena in the framework of Brans-Dicke theory, parametrised
by an additional parameter $\omega$ over and above those of General Theory of
Relativity (GTR), leading to very large values of $\omega$ ( GTR is the
$\omega\rightarrow\infty$ limit) is a superrior vindication of GTR than the
one done entirely within the framework of GTR. Speaking in the modern
language
of Renormalisation Group and Fixed Points, several RG-trajectories could, for
example, be flowing towards the same infra-red fixed point and in the
vicinity of that fixed point several flows could be practically
indistinguishable ( this is illustrated in fig. 1). a wider analysis of the
data would be in terms of the universality class to which the fixed point
belongs while an analysis based on a particular theory would correspond to
one of the many trajectories flowing into the fixed point.
\begin{figure}[htb]
\begin{center}
\mbox{\epsfig{file=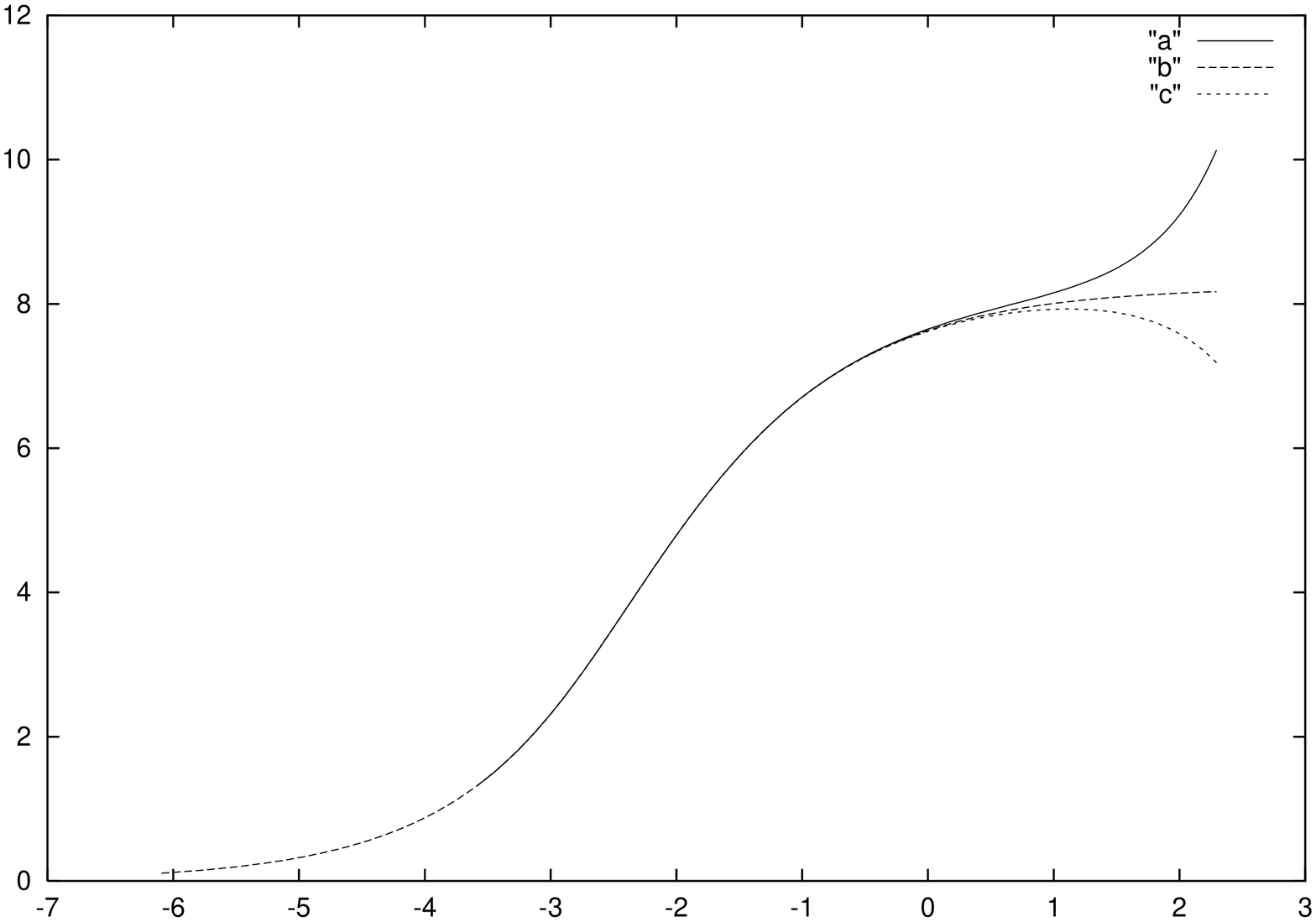,width=12truecm,height=8truecm,angle=-90}}
\caption{  Flows with an Infra-red Fixed Point }
\label{Fig 1.}
\end{center}
\end{figure}

Such a broader framework must necessarily possess the property of AF without 
which it would be impossible to make any credible statements about deep 
inelastic phenomena. Also, on theoretical grounds it is believed today that
only those theories possessing AF can be consistent \cite{hooft0}\footnote { In a
pragmatic sense one could argue that since we do not believe even theories
like QCD or the Standard Model to be valid at arbitrarily large energy
scales, the issue of whether a theory should possess AF is a moot one though
the presence of AF in the sense of a region where couplings are small and
get smaller with scale, would be important from a phenomenological point
of view. Even such a pragmatic view would perhaps admit that the "ultimate
theory" would possess AF.}.
While the relationship between AF and scaling is not straightforward, as
argued by Gross, non-AF theories are unlikely to lead to scaling.
From the works of Gross
and Coleman\cite{gross} it is well known that non-Abelian gauge fields are indispensable, 
at least in four dimensions, for realising AF. It is also by now well known 
that the matter content of gauge theories has important repercussions for the
survival of AF. For example, for SU(3) colour gauge theories, more than 16
families of quarks in the fundamental representation of the gauge group would
spoil AF. However, relatively less is known about the influence of scalar
fields on AF. 
	  In some specific cases, depending on the choice of the scalar 
representation
	  AF could be obtained by fine-tuning the scalar self-coupling , but did not occur
	  naturally as in QCD. To paraphrase G. t'Hooft: " In gauge
	  theories with fermions and scalars the situation is
	  complicated. If one imposes the condition that all beta
	  functions come out negative one usually finds that all
	  coupling constants should keep fixed ratios with each
	  other"\cite{spell}. However, in this paper we show that the
	  situation
	  need not be as restrictive as that.

In this paper we consider alternatives to QCD by enlarging the matter 
sector to include elementary scalar fields. We are using the phrase " matter
fields" in a generalised sense to include all fields that couple directly to
quarks or to gluons. One may argue that what we have considered here are
really alternatives to the Standard Model. We prefer to view our
attempts as alternatives to QCD because the symmetry structure of the
extended sector is $SU(2)_L\otimes SU(2)_R$. Of course inclusion of the
$U(1)$-sector of electroweak interactions eventually reduces the $SU(2)_R$
part of this symmetry.

We restrict our attention to cases where the additional scalars are colour 
singlets. This restriction can obviously be removed and an even larger class
of theories studied. We postpone such an analysis to the future. Of course, if the scalars were not colour singlets, even the one loop $\beta$-function for the
QCD coupling constant would change and it is doubtful whether the nice features
obtained here would survive. In fact it is believed that in QCD it will
be hard to add scalars (coupled to gluons) while keeping AF because the
quarks are in the fundamental representation and that that is the reason
why fundamental scalars that couple to the strong gauge group can not
exist\cite{spell}.

This choice of the colour singlet scalars was strongly motivated for us by the 
fact that the Gell-Mann-Levy linear $\sigma$-model with quarks substituted for 
nucleons has been found to provide a very reasonable description of the nucleon as well as of the strong interactions at finite temperature and density. While
the linear model was not asymptotically free and could therefore 
be considered only as
a low energy effective theory, the tantalising new question that
emerges is that of the AF of the alternative theories with a chiral multiplet 
of scalar fields.
It was established in \cite{pre} that such a model 
can indeed be AF in certain regions of the parameter space. What was however 
not clear at that time was the relation of this model to QCD itself; even the 
possibility that this theory is indistinguishable from QCD could not be ruled 
out.

In the present work we consider this issue at length by first establishing 
that there are indeed classes of such  theories that are asymptotically
free in all the additional couplings
and that AF is stable against inclusion of the electro-weak sector
\footnote{This is of course true only when the U(1) couplings are ignored.
Also, the Higgs-Yukawa and the Higgs self-coupling of the standard model
violate AF. These features are not altered in our alternative models.}, 
and that not all of them are
equivalent. We then proceed to confront these classes with existing 
experimental
data to see their viability. 
          Our analysis shows that for deep inelastic scattering 
          the leading
          behaviour of this theory in the ultra violet is identical to that of
          QCD. This is so as  the Yukawa coupling and the
          scalar self coupling go to zero faster than the QCD coupling. In fact,
          the numerical factors that arise are such
          as to make even the subleading QCD corrections to be marginally larger than
          the contributions of these extra couplings. This is important as some
          high energy experiments are already sensitive to these subleading
          corrections.

	 We further find that the usual strong interaction tests for QCD based 
          on the properties of quarks and gluons  will all go
	 through for this theory.
	 The other set of tests for the particle content of QCD 
	 include the
	 precision measurements like the R parameter and the  g-2 for the
	 muon. Here we will have extra contributions from our new scalar color
	singlet partons which couple to the photon. However , we find both
	these precision tests cannot confidently rule out our theory in favour 
of QCD.

A careful analysis of the flavour changing neutral currents of our model 
reveals more or less
uniform coupling to all flavours if significant flavour changing 
neutral currents are to be avoided. 
In contrast,extensions of the standard
model available in the literature like `two Higgs Doublet models'
(THDM),supersymmetric extensions of standard model like SSM,MSSM etc
generically predict dominant coupling to heavy flavours as well
as coupling to associated leptons.
Our model  naturally avoids coupling of the chiral scalars with
leptons.

The most important conclusions we have reached are 
that while the alternatives with massless 
or nearly massless chiral multiplets are ruled out by high precision 
Z-width data\cite{ztest}
(interestingly they are ruled out at present only by this one observable), 
those with suitably massive chiral multiplets 
can not be ruled out by current experimental 
data. We then propose a variety of experimental tests for such extended models
that have hitherto not been performed.

In such a theories, 
a distinctive signal will be the appearance of an excess of
four jet events in $e^+ e^-$ - collisions over what is to be expected 
from the standard model.
Such four jet signals arise as the massive scalars eventually decay into
$\bar q q$ pairs.

Though the ALEPH collaboration \cite{aleph1,rag,dorn} had reported seeing such an excess of
four jet events in $e^+ e^-$ collisions at $\sqrt s = $ 130,136,161 and
172 Gev respectively, they have subsequently attributed the excess to fluctuations
and do not see any excess in the later runs. Nevertheless, it is intertesting
that so many features of the additional four jet events that follow naturally
and in a parameter-free manner in our theories\cite{aleph} were reported by ALEPH in their
earlier claims. The disappearance of the ALEPH events does not repudiate our
model. It only raises the limit on the mass of the scalars which is also
consistent with other precision searches for additional heavy particles.
The present lower limit on the masses of such particles is around 100 GeV.

The precision tests of the standard model
based on the so called oblique parameters S,T\&U \cite{pesk,prob,hagi} 
also do not rule out the extensions considered here.
\section{The Model}
Our model is described by the lagrangian 
\begin{eqnarray}
{\cal L} = & - & \frac{1}{2} (\partial_\mu \sigma)^2 -\frac{1}{2} (\partial_\mu \vec\pi)^2
-\lambda^2 {(\sigma^2+\vec\pi^2-f_\pi^2)}^2\nonumber \\
& - & \overline\Psi_q\left[{\cal D}_\mu+
g_y(\sigma+i\gamma_5 \vec\tau\cdot\vec\pi)\right ]\Psi_q
-\frac{1}{4}G_{\mu\nu}^a G^{\mu\nu a}
\label{lagan}
\end{eqnarray}
where ${\cal D}_\mu=\partial_\mu -ig_3A_\mu^aT^a $ and $G_{\mu\nu}^a = \partial_\mu A_\nu^a - \partial_\nu A_\mu^a+ g_3 f_{abc} A_\mu^b A_\nu^c$. 
$A_\mu^a$ is the gluon field and $T^a$ the SU(3) generator in the fundamental representation.
$g_y$, $g_3$ and  $\lambda$ are the Yukawa, QCD, and scalar self- couplings respectively.
 $\overline\Psi_q$ is the quark field. 

The $\sigma,\vec\pi$ in eqn (1) should not be confused with the chiral
multiplet occurring in the low lying spectrum of strongly interacting
particles
containing the pion($m_{\pi}\simeq 140 MeV$) though they have the same
quantum numbers. While the latter are bound states the $\sigma,\vec\pi$ of
eqn (1) are elementary, to be thought of as additional "partons" of the
model. The important property of eqn (1) is that it is invariant under the
{\bf Global} $SU(2)_L\otimes SU(2)_R$ transformations that affect the quark
fields and chiral multiplet. Introducing
\beeq
U = \sigma+ i\gamma_5\vec\tau\cdot\vec\pi
\eneq
these transformations are given by
\beeqar
\psi_L^{'} &=& g_L\psi_L\nonumber\\
\psi_R^{'} &=& g_R\psi_L\nonumber\\
U^{'}&=& g_LU g_R^{\dag}
\eneqar
It is further assumed that in the extended sector there is no spontaneous
breaking of $SU(2)_L\otimes SU(2)_R$.

As shown in \cite{ztest}, a model described by
 eqn(1) 
 automatically implies that the chiral multiplet couples to the
 electroweak gauge bosons. 
Representing the chiral multiplet as a complex doublet
\beeq
\Phi_{ch}^{T} = (\sigma+i\pi_0,i\sqrt{2}\pi_{-})
\eneq
The coupling of the chiral multiplet to the electroweak gauge bosons is
given by the minimal coupling
\beeq
{\cal L}_{gauge,chiral} = {1\over 2}|(\partial_\mu\Phi^{ch}-i{g\over 2}
\vec\tau\cdot \vec A_{\mu}\Phi^{ch}-i{g^{\prime}\over 2}YB_{\mu}\Phi^{ch})|^2
\eneq
with $Y=-1$ and $g,g^{'}$ be the $SU(2)_L$ and $U(1)$ couplings respectively.
For example,the linear couplings that result are
\beeqar
{\cal L}_{lin}^{neut}&=& e(A_{\mu}-{\gamma\over 2} Z_{\mu})(\vec\pi\times
\partial_\mu\vec\pi)_3
-{e\over 2cs}Z_\mu(\pi_0\partial_\mu\sigma-\sigma\partial_\mu\pi_0)
\nonumber\\
{\cal L}_{lin}^{char}&=& {g\over 2}W_-^{\mu}[(\pi_+\partial_\mu\sigma-\sigma
\partial_\mu\pi_+)+i(\pi_+\partial_\mu\pi_0-\pi_0\partial_\mu\pi_+)]\nonumber\\
& &+c.c
\eneqar
where $\gamma = (1-2s^2)/cs)$ and $c,s$ are $\cos\theta_W,\sin\theta_W$
respectively. Likewise, the quadratic gauge couplings are given by
\beeqar
{\cal L}_{quad}&=&{e^2\over 8s^2c^2}Z_\mu Z^\mu(\sigma^2+\pi_0^2)+
g^2/2~W_\mu^+W^{-\mu}\pi_+\pi_-\nonumber\\
&+i&{eg\over 4sc}[Z^\mu W_\mu^-(\sigma+i\pi_0)\pi_+-Z^\mu W_\mu^
+(\sigma-i\pi_0)\pi_-]\nonumber\\
&+&g^2/4~W_\mu^-W^{+\mu}(\sigma^2+\pi_0^2)\nonumber\\
&+&e^2(A_\mu A^\mu+\gamma^2/4~ Z_\mu Z^\mu-\gamma A_\mu Z^\mu)\pi_
+\pi_-\nonumber\\
&+&ieg/2(A_\mu-\gamma/2 Z_\mu)W^{-\mu}(\sigma+i\pi_0)\pi_+\nonumber\\
&-&ieg/2(A_\mu-\gamma/2 Z_\mu)W^{+\mu}(\sigma-i\pi_0)\pi_-
\eneqar
\subsection{Higgs Sector}
Since the additional multiplet has to be massive,it is economical to have
their masses generated by the Higgs mechanism. The simplest way to do this is
by introducing the coupling
\begin{equation}
{\cal L} = {\bar\lambda \over 2}|\Phi|^2 (\sigma^2 + \vec\pi ^2)
\end{equation}
Then $\lambda$ is fixed to be
\begin{equation}
\bar\lambda = {m_c^2 \over v^2}
\end{equation}
where $m_c$ is the chiral multiplet mass and $v$ the vacuum expectation value
of the Higgs field. Now, in the broken phase, the physical Higgs field couples
to $\sigma $ and $\vec\pi$ according to
\begin{equation}
{\cal L} = {m_c^2\over v}H (\sigma^2 + \vec\pi ^2)
\end{equation}
Note that both eqn (1) and eqn (10) are invariant under $SU(2)_L\otimes
SU(2)_R$ of eqn (3).
If the Higgs is not too light but lighter than the top quark, this channel
would be the most dominant mode for Higgs to decay and would open a new window
into Higgs search. 
This additional coupling to Higgs raises new issues;for example,
the renormalisation group eqns for the flow of both the Yukawa coupling $g_y$
and the scalar self coupling $\lambda$ will now change due to the additional
coupling $\bar\lambda$. More importantly, the renormalisation group flow for
the Higgs coupling parameters of the standard model will change due to the 
coupling of the Higgs to the chiral multiplet considered here. These issues
will be addressed elsewhere. It also raises the issue of radiative mixing between the chiral 
multiplet and Higgs particle, which by design is absent at tree level. This
is discussed in detail later in the text.
\subsection {Unitarity Constraints}
One of the key properties of the standard model is the unitarity of all
scattering amplitudes. The standard model Higgs 
kills the bad high-energy
behaviour of processes like $WW \rightarrow WW$ and $WW \rightarrow
\bar f f$ where $f$ is any fermion. Introduction of new bosons
into
the theory that couple to fermions and gauge bosons should 
not spoil this . The necessary and
sufficient conditions for this are \cite {haber}
\beeqar
\Sigma_i g_{h^0_i V V}^2 & = & g_{hVV}^2\nonumber\\
\Sigma_i g_{h^0_i V V} \cdot g_{h^0_i f \bar f}& = & g_{hVV}\cdot
g_{hf\bar f}
\eneqar
where h is the standard model Higgs and $h_i^0$ are all the neutral
scalar fields of the theory including the analog of h. In our theory,
since there is no SSB in the chiral - multiplet sector, no $\chi VV $
coupling is introduced and the above conditions are trivially
satisfied.
\section{The RNG Flows}
The one-loop $\beta$ function for the QCD coupling, $\alpha$, is
\begin{equation}
\frac{\partial \alpha}{\partial  t} = - \left ( \frac{33 - 2 N_F}{3}\right ) \frac{\alpha^2}{8\pi^2} \hspace{5em} \left ( g_3^2=\alpha \right )
\label{dadt}
\end{equation}
where $t=\ln (p/\mu)$.\\
To 
this order the $\beta$-function
for the QCD coupling does not receive any contribution from $g_y$, 
$\lambda$, $g$ or $g^{\prime}$ as neither the chiral multiplet nor the 
electro-weak gauge bosons couple directly to gluons.

The situation with the beta functions for $g_y,\lambda$ is , however, more
intricate. If the $U(1)$-coupling $g^{\prime}$ is large, radiative corrections
will spoil the $SU(2)_L\times SU(2)_R$ structure. Instead of a single Yukawa 
coupling as in eqn(1), one would have
\beeq
{\cal L}^{broken}_{Yukawa} = g_{y1}(\bar u_L \bar d_L)\Phi u_R +
                             g_{y2}(\bar u_L \bar d_L)\Phi^{\prime} d_R +h.c
\eneq
where $\Phi^{\prime} = C\Phi^{*}$ is the charge-conjugate of $\Phi$. We will
show shortly that the $U(1)$-coupling can be neglected in the one-loop
analysis and one can continue to use the form as in eqn(1). Before doing that
we derive the full beta functions(one-loop). In the minimal subtraction
scheme, the renormalisation constants relevant for this discussion are given by:
\beeqar
Z^{wf}_{\sigma} & = &  1 + {1\over 8\pi^2\epsilon}[-6N_g^{\prime}( {g_{y1}}^2+
                           {g_{y2}}^2)+{3g^2 +{g^{\prime}}^2\over 2}]\nonumber\\
Z^{wf}_{u_L} & = &  1 + {1\over 8\pi^2\epsilon}[{\bf- ({g_{y1}}^2+
                           {g_{y2}}^2)}-{27g^2 +{g^{\prime}}^2\over 36}-{4g_3^2\over 3}]\nonumber\\
Z^{wf}_{u_R} & = &  1 + {1\over 8\pi^2\epsilon}[{\bf-2 {g_{y1}}^2}
                           -{4{g^{\prime}}^2\over 9}-{4g_3^2\over 3}]\nonumber\\
Z^{wf}_{d_L} & = &  1 + {1\over 8\pi^2\epsilon}[{\bf-( {g_{y1}}^2+
                           {g_{y2}}^2)}-{27g^2 +{g^{\prime}}^2\over 36}-{4g_3^2\over 3}]\nonumber\\
Z^{wf}_{d_R} & = &  1 + {1\over 8\pi^2\epsilon}[{\bf
                           -2{g_{y2}}^2}-{{g^{\prime}}^2\over 9}-{4g_3^2\over 3}]\nonumber\\
Z^{v}_{g_{y1}}      & = &  1 + {1\over 8\pi^2\epsilon}[ {\bf 2{g_{y2}}^2}+
                            {3g^2\over 4} + {25{g^{\prime}}^2\over 36}+{16g_3^2\over 3}]\nonumber\\
Z^{v}_{g_{y2}}      & = &  1 + {1\over 8\pi^2\epsilon}[{\bf 2{g_{y1}}^2}+
                            {3g^2\over 4} + {{g^{\prime}}^2\over 36}+{16g_3^2\over 3}]
\eneqar
Here 'wf' and 'v' refer to the wave-function and vertex renormalisations
respectively. $N_g^{\prime}$ is the effective number of flavours to which the chiral multiplet
couples(see the discussion on flavour changing neutral
currents in sec.5.1 which forces equal coupling of the chiral multiplet to all flavours). The resulting beta - functions are
\beeqar 
{dg_{y1}\over dt} & = & -{g_{y1}\over 8\pi^2}[-3N_g^{\prime}(g_{y1}^2+g_{y2}^2)
                    +{\bf {3\over 2}(g_{y2}^2-g_{y1}^2)}+
                     {27g^2+17{g^{\prime}}^2\over 24}+4g_3^2]\nonumber\\
{dg_{y2}\over dt} & = & -{g_{y2}\over 8\pi^2}[-3N_g^{\prime}(g_{y1}^2+g_{y2}^2)
                    -{\bf {3\over 2}(g_{y2}^2-g_{y1}^2)}+
                     {27g^2+5{g^{\prime}}^2\over 24}+4g_3^2]\nonumber\\
\eneqar
In these equations, the terms in bold are the ones arising out of loops with
$M_{\chi}$ as the largest mass (except the cases where the one of the internal
fermion lines is that of top quark). The significance of these terms will be
clarified shortly. Already at this stage it is clear that the $U(1)$-coupling
$g^{\prime}$ has a negligible influence. This follows from the fact that
$g^2 \simeq {g^{\prime}}^2$ and that $g^2$ is already about a fourth of $g_3^2$
(say, near the scale of the Z-mass; at lower scales, due to the non-asymptotically
free nature of $g^{\prime}$, the importance of it becomes even less). But this
would be a little hasty though correct observation. To see this, let us
introduce, following \cite{schre}, $\rho_1 =g_{y1}^2/\alpha, \rho_2 = g_{y2}^2/\alpha,
 \xi = g^2/\alpha$ and $\xi^{\prime} = {g^{\prime}}^2/\alpha$, it is easy to obtain
\beeqar
{\partial \rho_1 \over \partial\alpha} & = & -{\rho_1\over A\alpha}[6N_g^{\prime}(\rho_1
+\rho_2)-{\bf 3(\rho_2-\rho_1)}-(8+{17\over 12}\xi^{\prime}+{9\over 4}\xi-A)]\nonumber\\ 
{\partial \rho_2 \over \partial\alpha} & = & -{\rho_2\over A\alpha}[6N_g^{\prime}(\rho_1
+\rho_2)+{\bf 3(\rho_2-\rho_1)}-(8+{5\over 12}\xi^{\prime}+{9\over 4}\xi-A)] 
\eneqar
where $A = 11 - 2N_F/3$. Thus we see that the contributions due to $\xi^{\prime}$
should really be negligible in comparison to $A-8-9\xi/4$ in order that we can
ignore the $U(1)$-couplings as it is the sign of $A-8-9\xi/4-c\xi^{\prime}$
in eqns(14) that determines whether the flows are asymptotically free or not. It
is also significant for this discussion that all the gauge couplings $g_3,g,g^{\prime}$
enhance the tendency towards asymptotic freedom. For $N_F=6, 5$, $A-8=1,1/3$ respectively
and one can have AF regions for $\rho_1,\rho_2$ irrespective of the values of
$\xi,\xi^{\prime}$. At a scale $\simeq M_Z$ where $A-8 = -1/3$, $9\xi/4\simeq 0.6$ and $17\xi^{\prime}
\simeq 0.12$ and thus neglecting the $\xi^{\prime}$ contribution leads to an
error of about 10\%.In the eqn for the average of $\rho_1,\rho_2$ this error is
in fact only about 8\%. It should however be kept in mind that eventually the
$U(1)$-coupling grows in the asymptotia because of the lack of asymptotic freedom in this coupling. 
But
in practical terms this coupling begins to equal the other couplings only around $\simeq 10^{14} GeV$
or so, which is a scale far beyond the scope of our model. We therefore emphasise that
our discussion of asymptotic freedom is within the natural framework where asymptotia
means a scale much larger than all the natural scales of the theory and for
our model should be thought of as in the range 100-1000 GeV.

It should be observed that if we neglect the $\xi^{\prime}$ contribution,
a consistent solution to eqn(14) is $\rho_1=\rho_2$. In the Wilson RNG spirit
we take this branch as defining our theory. Then the terms in boldface in these eqns 
vanish with the consequence that as we cross the mass threshold at $M_{\chi}$,
 the flow eqns of the theory are unmodified. We shall return to a discussion of the effects
of various mass thresholds shortly.

Thus we can return to analysing the theory in terms of a single $g_y$ 
and in terms of $\rho = g_y^2/\alpha$, we have
\beeq
{\partial \rho \over \partial\alpha} = -{\rho\over A\alpha}[12N_g^{\prime}\rho
-(8+{9\over 4}\xi-A)] 
\eneq
As it stands, the above eqn can not be integrated analytically because $\xi$
is a function of $\alpha$. It is however illuminating to consider eqn(15) after
neglecting the $\xi$-dependent term, as in that case the eqn is analytically
soluble and possesses qualitatively similar features as the full equation. Thus
the equation we consider is
\beeq
{\partial \rho \over \partial\alpha} = -{\rho\over A\alpha}[12N_g^{\prime}\rho
-8+A)] 
\eneq
By defining$\tilde\rho = N_g^{\prime}\rho$, we can rewrite this as
\beeq
{\partial \tilde\rho \over \partial\alpha} = -{\tilde\rho\over A\alpha}[12\tilde\rho
-8+A)] 
\eneq 
For the $N_F=6$ case where $A=7$, calling $\tilde\rho_c=1/12$,
 there are two regimes:

 {\bf The Region $0< \tilde\rho<\tilde\rho_c$}: 
 
 In this case $\partial \tilde\rho/\partial \alpha>0$. 
This implies that $\tilde\rho$ decreases as $\alpha$  decreases, 
that is, $\tilde\rho$ will decrease with increasing momentum scale. We can integrate the $\tilde\rho$ equation to get
\begin{equation}
\tilde\rho=\frac{\alpha^{1/7} K}{(1+12 \alpha^{1/7} K)}
\end{equation}
where $K$ is a positive integration constant that is set by initial data on 
$\alpha$ and $g_y^2$. Specifically
\beeq
K = {\tilde\rho_0\alpha_0^{-1/7}\over 1-12\tilde\rho_0}
\eneq
It is therefore clear that there is a whole family of
solutions corresponding to different K's. 
Deep in the ultraviolet when $\alpha\rightarrow 0$ $$\tilde\rho\sim K\alpha^{1/7}$$
 which further implies that in the ultra violet $$N_g^` g_y^2 \sim K\alpha^{8/7}$$
 This means that $g_y^2$ is asymptotically free and vanishes faster than 
 $\alpha$. Therefore, the leading behaviour of this theory in the ultraviolet is
 given by that of standard QCD , with the yukawa coupling contributing only in 
sub leading order. \par
\begin{figure}[htb]
\begin{center}
\mbox{\epsfig{file=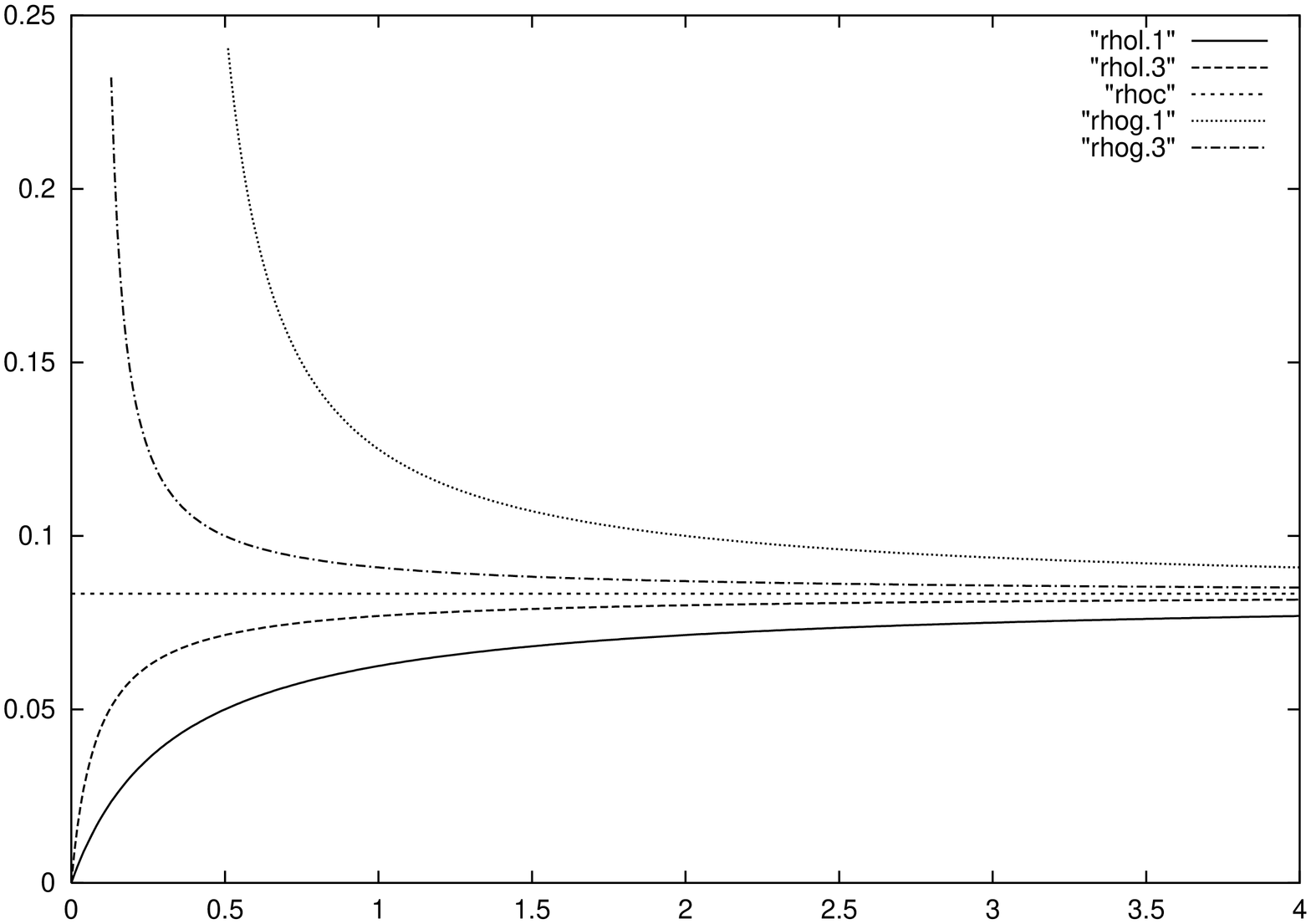,width=12truecm,height=8truecm,angle=-90}}
\caption{ The renormalisation group flows for rho vs alpha.}
\label{Fig 2.}
\end{center}
\end{figure}
The region $\tilde\rho > 1/12 $ is also of interest, though from a different
physical perspective. However, the theory is not AF in this region. Therefore
 we shall not consider it anymore in this paper. 

The above analysis is valid for $q^2 \ge m_t^2$. For the region
$m_b \le q \le m_t$, the relevant behaviours are : $\tilde\rho_c = 1/36,
\tilde\rho \simeq K\alpha ^{1/23}$. 
For $N_F \le
4$ this approximate eqn does not admit any AF regime. But by the time the effective $N_F$ reaches 4,
$\alpha\simeq 2.74$ and the one-loop analysis need not be trusted.

Now let us return to the more precise eqn(15). As mentioned earlier, the explicit dependence of $\xi$
makes it difficult to analyse this equation analytically. However, the precise
form of $\xi$ can be obtained by integrating the flow equation for $\xi$. The
flow eqn for the $SU(2)_L$ coupling constant is given by \cite{cheng}
\beeq
{dg_2\over dt} = {g_2^3\over 16\pi^2}[-{11\over 3}t_2(V)
{4\over 3}+n_Ft_2(F)+{1\over 3}n_St_2(S)]
\eneq
where $n_F, n_S$ are the number of fermion and (complex) scalar multiplets 
in the fundamental representation of $SU(2)_L$. In the standard model as also in our model, if $N_g$
is the number of quark-lepton generations, $n_F=2N_g$. However, it is more instructive to
separate the lepton and quark contributions as all the lepton thresholds would
be contributing, but not all the quark thresholds. Then, $n_F={3\over 2}+{3N_F\over 4}$ where $N_F$ as before denotes the 
number of effective quark flavours. Again, in the standard model, $n_S$ is 1
(the standard Higgs doublet) while in our model it is $1+N_{\chi}$ where
$N_{\chi}$ is the effective number of chiral multiplets. Further, $t_2(V) =2$ and $t_2(f)=t_2(S)=1/2$
for $SU(2)$. Thus the relevant
flow equation for our model is
\beeq
{dg_2\over dt} = {g_2^3\over 16\pi^2}[-{37\over 6}
+{N_F\over 2}+{N_{\chi}\over 6}]
\eneq
thus
\beeq
{d\xi\over dt}= -{A_L\over 8\pi^2}\xi^2
\eneq
where $A_L = {37\over 6} -{N_F\over 2}-{N_{\chi}\over 6}$. On the other hand
\beeq
{d\alpha\over dt}= -{A\over 8\pi^2}\alpha^2
\eneq
The solutions to these equations are, respectively,
\beeqar
{1\over \alpha} &=& {A\over 8\pi^2}t + const\nonumber\\
{1\over \xi} &=& {A_L\over 8\pi^2}t + const
\eneqar
hence the quantity ${A\over \xi}-{A_L\over \alpha}$ is a RNG invariant. As a
consequence we have
\beeq
{A\over \xi}-{A_L\over \alpha } = {A\over \xi_0}-{A_L\over \alpha_0 } = {\cal A}(N_F,N_{\chi}, \xi_0, \alpha_0)
\eneq
This can be used to solve for $\xi$ as a function of $\alpha$:
\beeq
\xi = {A(N_F)\over A_L(N_F, N_{\chi})+{\cal A}\alpha}
\eneq
Substituting this in eqn (6) one gets
\beeq
{\partial \rho \over \partial\alpha} = -{\rho\over A\alpha}[12N_g^{\prime}\rho
-(8+{9\over 4}{A(N_F)\over A_L(N_F, N_{\chi})+{\cal A}} -A)] 
\eneq
This can be integrated numerically. But already some trends can be gleaned 
from this eqn;in the absence of the $\xi$ term, $\rho_c$ would be given by
$\rho_c = (8-A)/(12N_g^{\prime})$. But $\xi$ changes it to $\rho_c =
(8+{9\xi\over 4}-A)/(12N_g^{\prime})$. In the regime where $N_F=5, N_{\chi}=1$
(scales in the range 90 GeV-175 GeV), the $\xi$ contribution triples the value of
$\rho_c$! This is clearly of great relevance to the phenomenology of this model.
An accurate numerical solution of these eqns will be provided elsewhere.

\subsection{Threshold Effects}
We have treated the flow equations without regard to the various mass thresholds. A
precise treatment of this is very complicated and is not warranted right now.
However, the qualitative trends can be analysed. As one comes down from deep
asymptotia, the first mass threshold to be encountered(in our model) is the
top mass threshold at $\simeq 175 GeV$. Below this(actually much below this, but we shall
not go into such finer details), the effective $N_F=5$ and it remains at this value 
till we reach the bottom threshold at 5 GeV where 
it jumps to 4. In  regions where $N_F\le 4$ one loop results need not be reliable. 
As one goes below the threshold at $M_Z$, 
the $\xi$ contributions to the flow equations drop out. Of course, the 
$\xi^{\prime}$
contributions arising out of the massless photon exchanges are
always present, but due to the combined effects of the asymptotically free nature
of $\alpha$(which therefore keeps increasing towards the infrared) and the non-asymptotically
free nature of $g^{\prime}$(which therefore keeps decreasing towards the infrared)
, $\xi^{\prime}$ actually decreases very rapidly. As we have already remarked,the threshold
at $M_{\chi}$ has no effect. This is a very important property of the model
which allows the flow eqns derived above both for the scenario in which the chiral multiplet
is light mass as well as the case where they are considered to be at least
as massive as $M_Z/2$. Thus for $N_F\le 4$, our one-loop flow 
eqn reduces to the approximate form of eqn(16) which has 
no asymptotically free solutions. 

For the purposes of our paper (in the context of the discussion on
$g - 2$ for muons) it is important that regions where QCD is
perturbatively treatable, our theory is too. 
AF for the scale $ > m_b $ implies $\rho \le 1/36$.
Thus at $q = m_b$, the Yukawa couplings are small. Now the lack of AF for
$q < m_b$ has the desired effect of making $\rho$ even smaller as we go
to
smaller energy scales. Thus $g_y$ is perturbatively
treatable wherever QCD is. 

\subsection{Scalar Self-coupling}
So far we have not addressed the question of the RNG flows of the self-coupling
$\lambda$. We leave all details of the calculation and merely present the 
full $\beta$-function for $\lambda$ in our model: 
\beeq
{d\lambda\over dt} = {1\over 8\pi^2}[{\bf 2\lambda^2} + 24N_g^{\prime}\lambda
(g_y^2-{3g^2\over 8}-{{g^{\prime}}^2\over 8})-144N_g^{\prime}g_y^4+
({27g^4\over 8}+{9g^2{g^{\prime}}^2\over 4}+{9{g^{\prime}}^4\over 8})]
\eneq
It is again clear that the $U(1)$-coupling can be neglected. Again we can
introduce the ratio $R=\lambda/g_y^2$ following \cite{schre} and convert this flow equation
to
\beeq
[12N_g^{\prime}\rho -(8+{9\xi\over 4}-A)]{dR\over d\rho}=
{\bf 2R^2}+12N_g^{\prime}R+{8R\over \rho}-144N_g^{\prime}
+({9\over 4}-9N_g^{\prime}) {\xi\over \rho}R+{27\xi^2\over 8\rho^2}
\eneq
As before, let us first consider an approximation to this eqn by dropping
the $\xi$-dependent terms:
\beeq
[12N_g^{\prime}\rho -8+A]{dR\over d\rho}=
{\bf 2R^2}+12N_g^{\prime}R+{8R\over \rho}-144N_g^{\prime}
\eneq
This is the same type of eqn that was considered in \cite{schre}; it depends only 
on $\rho$ and can be easily solved analytically.
It is found that only on a single trajectory in the $[R, \rho]$ parameter 
space, that is the so called invariant
line~\cite{ross}-\cite{har}, the behaviour of $R$ for the regime
$N_g^` \rho <1/12$ is such that $R\rightarrow 0$ in the ultraviolet. 
It follows further that $\lambda\rightarrow 12N_g^{\prime}\rho$ in the extreme ultraviolet 
even faster than $g_y^2$. 
Note that such a possibility is not available with eqn(21).

In figs. 3 and 4 we have shown the flow diagrams for R according to eqn
(30) for the case $N-g^{'}=3(A=7)$. Of particular importance is the
function $R_+(\rho)$ which is the positive root of
\beeq
R^2+18 R + 4 R/{\rho} - 216 = 0
\eneq
Fig. 3 shows $R_+(\rho)$ as well as the unique trajectory ( the
invariant line ) on which R is regular everywhere. Every point on this
line flows to $R=0$ in the UV as $R=48\rho$ and approaches $R_{\infty} = -9
+\sqrt{279}$ in the IR. For points such that $R > R_+$, ${dR\over d\rho}$
is positive. The point $ \rho = 1/36$ at which the value of R on the
invariant line is $R_c = -81 + \sqrt{6777}$ is shown by the cross. The
typical flows are shown in fig. 4. The curve above the invariant line
but to the right of $\rho = 1/36$ is such that all points on it flow to
$R = \infty$ in the IR, but approach the invariant line in the UV. The
flow to the right of $\rho = 1/36 $ but below the invariant line is such
that all points on it flow to $R = 0$ in the IR at some finite scale and
eventually to negative R, but converge to the invariant line in UV.
Likewise, the flow to the left of $\rho = 1/36$ which is above the
invariant line is such that all points on it move to $R_{\infty}$ in the
IR but to $R=\infty$ at a finite scale in the UV. The flow to the left
of $\rho = 1/36$ below the invariant line is such that it flows to
$R_{\infty}$ in the IR but goes to $R = 0$ at a finite scale in the UV
and eventually to negative values of R.

\begin{figure}[htb]
\begin{center}
\mbox{\epsfig{file=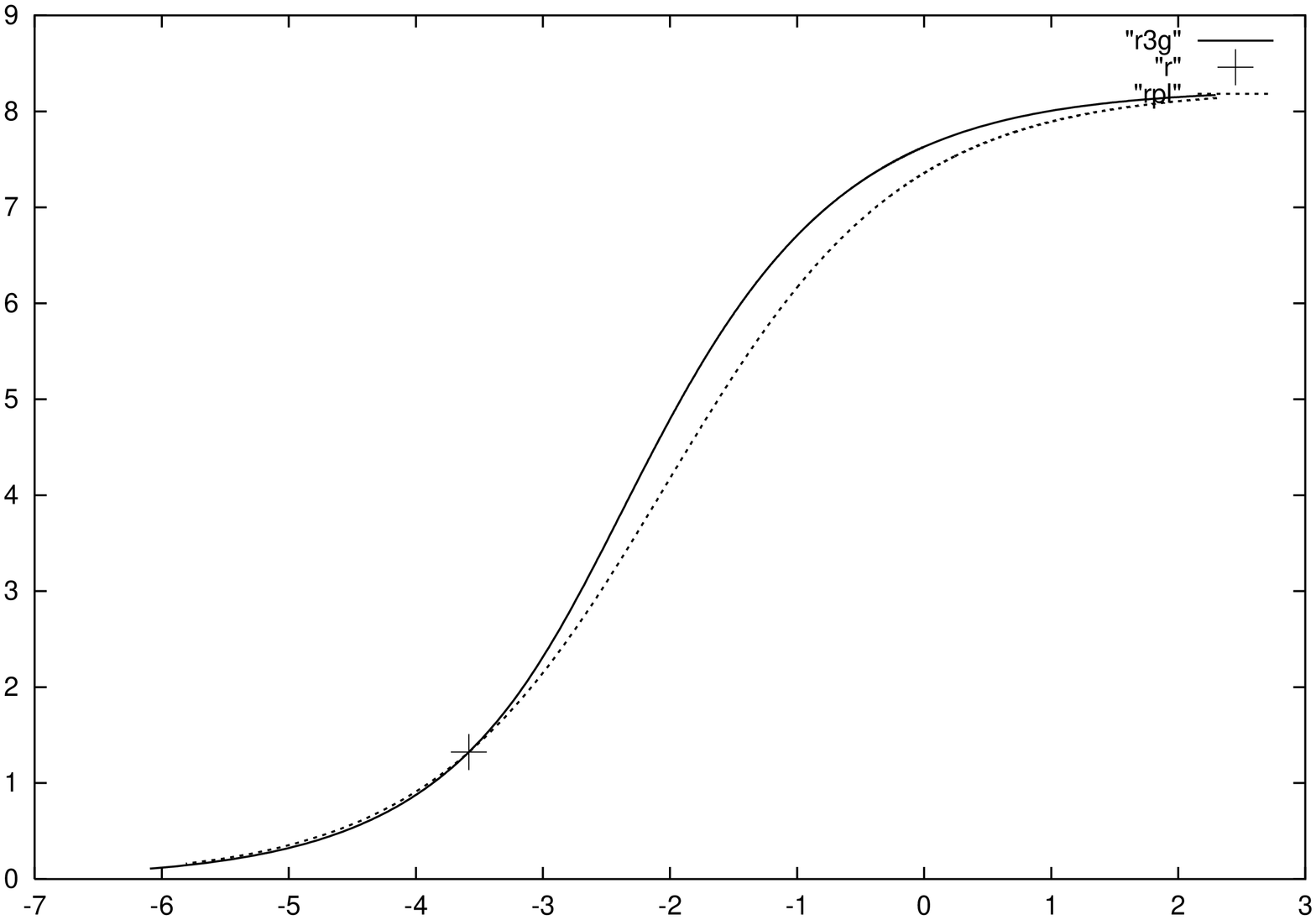,width=12truecm,height=8truecm,angle=-90}}
\caption{ The Invariant-line and the curve $R_{+}(\rho)$}
\label{Fig 3.}
\end{center}
\end{figure}
\begin{figure}[htb]
\begin{center}
\mbox{\epsfig{file=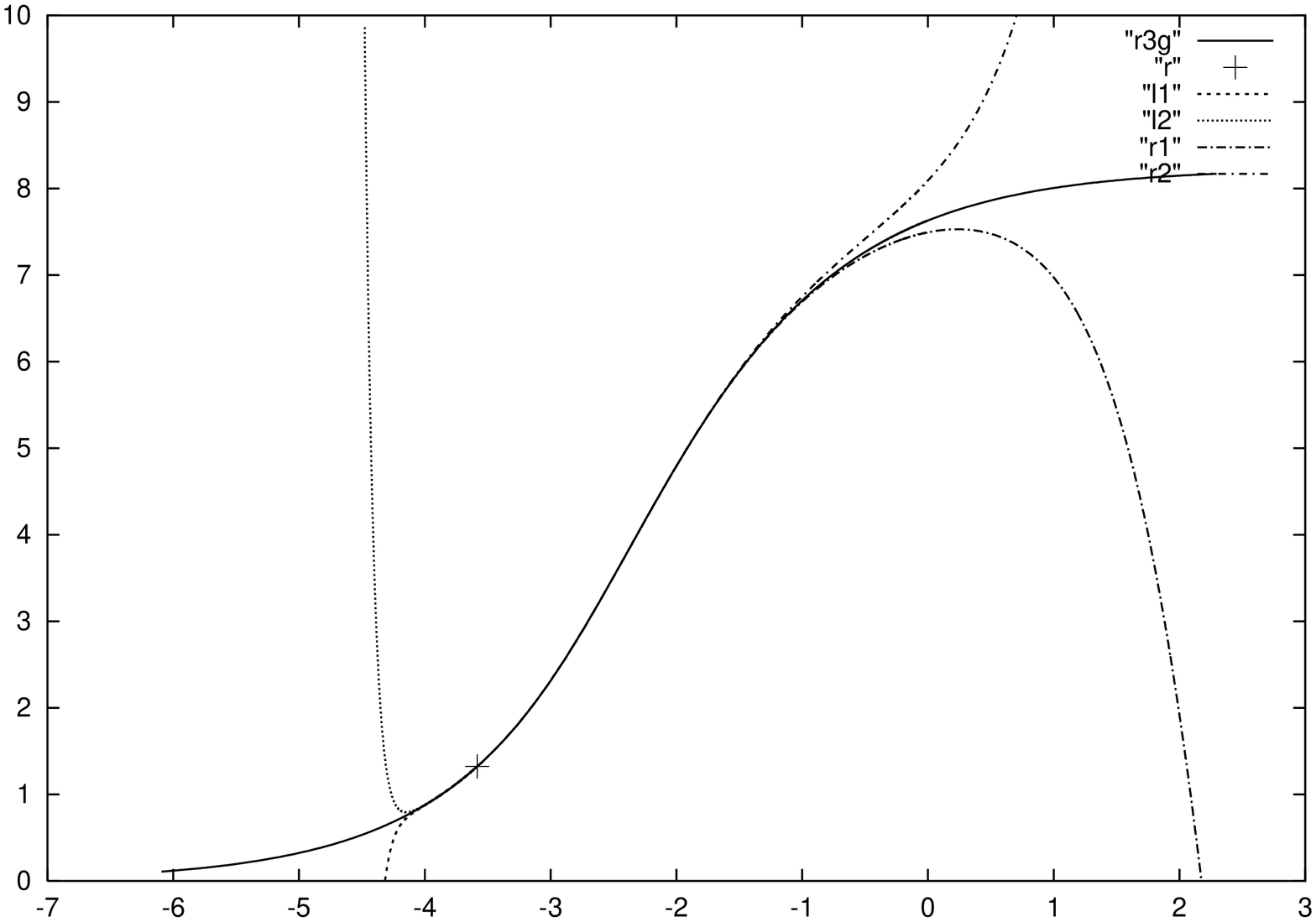,width=12truecm,height=8truecm,angle=-90}}
\caption{ The Renormalisation Group Flows for R vs rho.}
\label{Fig 4.}
\end{center}
\end{figure}
On the other hand, the structure of the eqn(29) shows that there can be no solution
for $R$ that is regular as $\rho\rightarrow 0$. But there is no reason for $R$
to be regular for $\lambda$ to be asymptotically free because $\lambda$ could
be vanishing slower than $g_y^2$ and yet be asymptotically free. In fact an inspection 
of eqn(29) shows that
as $\rho\rightarrow 0$ $R$ must approach $c/\rho$ where c satisfies
\beeq
2c^2+8c-{27\over 4}{A\over A_L}c+{27\over 8}{A^2\over A_L^2}-(8-A+{9A\over 4A_L})=0
\eneq
Deep in asymptotia, $N_F=6$ and $N_{\chi}=1$ and consequently $A = 7$ and $A_L =3$
and this eqn becomes
\beeq
2c^2-14c+{147\over 8} = 0
\eneq
with solutions $c = {21\over 4}, {7\over 4}$. The nature of these solutions
is completely different from the solutions we got without $\xi$. It should be
emphasised that $R=c/\rho$ means $\lambda = c\alpha$.

Thus we have classes of theories that are not only AF in all their couplings,
 but become increasingly indistinguishable from QCD at high energies. 
As far as AF is concerned 
one loop analysis is stable against higher loop corrections 
\cite{gross}. 
Since these classes of theories are AF, they are all candidates for a consistent theory of strong interactions.
\begin{figure}[htb]
\begin{center}
\mbox{\epsfig{file=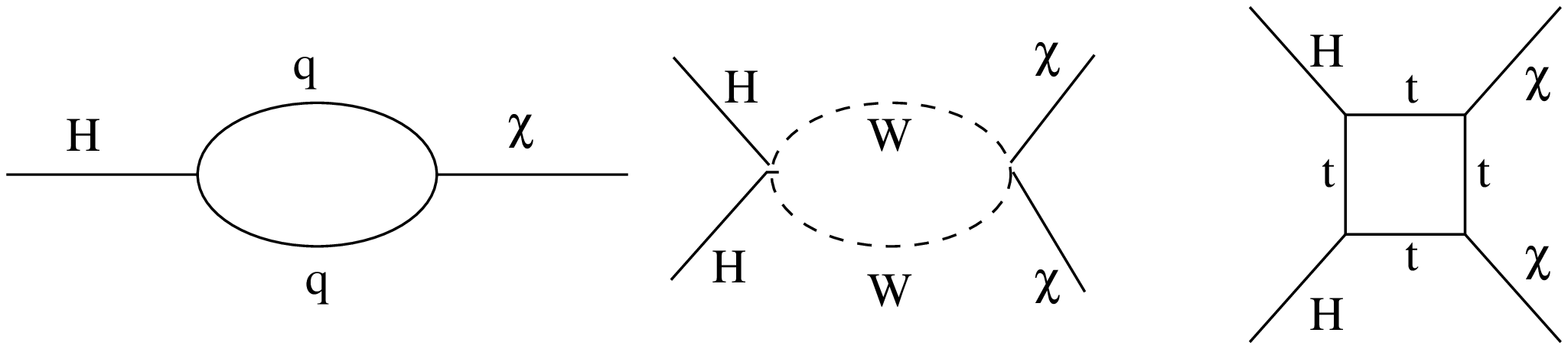,width=12truecm,height=4truecm,angle=-90}}
\caption{ Typical diagrams contributing to Higgs-Chiral Scalar mixing}
\label{Fig 5.}
\end{center}
\end{figure}
\subsection{Higgs- Chiral Scalar Mixing}
Since both the chiral multiplet and the standard model Higgs doublet
couple directly to quarks and electroweak gauge bosons, quantum
fluctuations generically lead to a mixing between them. The typical loop
diagrams that contribute to such mixing is shown in fig. 5. These
diagrams are also divergent. Thus renormalisability of the model would
require the most general Higgs-Chiral Scalar coupling and not just the
coupling in eqn (8). It is also clear that the mixing generated by these
diagrams reduces $SU(2)_L\otimes SU(2)_R$ to $SU(2)_L\otimes U(1)$. In
current literature one usually finds such couplings under the assumtion
of some discrete symmetries. Such discrete symmetries which in our case
are present at tree level, are broken by quantum corrections. The most
general such potential without assuming any discrete symmetries is
\beeqar
V &=& \mu_1^2 \Phi_1^{\dag}\Phi_1 + \mu_2^2 \Phi_2^{\dag}\Phi_2 +
(\mu_3\Phi_1^{\dag}\Phi_2 +h.c)+ \lambda_1(\Phi_1^{\dag}\Phi_1)^2+
\lambda_2(\Phi_2^{\dag}\Phi_2)^2+\nonumber\\
&&\lambda_3(\Phi_1^{\dag}\Phi_1)(\Phi_2^ {\dag}\Phi_2)+(\lambda_4
(\Phi_1^{\dag}\Phi_2)^2+\lambda_5(\Phi_2^{\dag}\Phi_1)^2
+ \lambda_6(\Phi_1^{\dag}\Phi_1)(\Phi_2^{\dag}\Phi_1) +\nonumber\\
&&  \lambda_7(\Phi_2^{\dag}\Phi_2)(\Phi_2^{\dag}\Phi_1)
+\lambda_8(\Phi_1^{\dag}\Phi_1)(\Phi_1^{\dag}\Phi_2)
+\lambda_9(\Phi_2^{\dag}\Phi_2)(\Phi_1^{\dag}\Phi_2)+h.c)
\eneqar
Our model at tree level is assumed to be such that only $\mu_1,\mu_2,
\lambda_1,\lambda_2,\lambda_3$ are non-vanishing. When quantum
fluctuations are taken into account, all these parameters become
scale dependent and one can not consistently put them to zero at all
scales. But the scale evolution of these parameters could be such that
for all practical purposes the parameters chosen to be zero at tree
level remain very small for all scales of interest. To establish this 
completely satisfactorily requires a lot of work. Instead, we shall show
the reasonableness of this by estimating the sizes of various diagrams
after they have been renormalised to keep the renormalised values of
these parameters at zero at some scale.

In the self energy type of graphs of fig. 5, the dominant contribution
comes from the top quark loop. The Higgs- top coupling is taken to be
$\eta\simeq 3/4$. Taking $\alpha_s = \alpha/(4\pi)\simeq 0.1 $ and
$\rho=1/36$, one gets $g_y\simeq 1/5.5$. The self-energy graph at top
mass scale can thus be estimated to be $\simeq {\eta g_y\over
8\pi^2}m_t^2\simeq m_t^2/500$. Thus the expected magnitude of
$\mu_3^2\simeq m_t^2/500$ which should be compared with $\mu_1^2\simeq
m_H^2,\mu_2^2\simeq M_{\chi}^2$. Thus the mixing effect here is indeed
small. The mixing terms that are quartic in fields can be of the type
$H^3\chi,H^2\chi^2,$ or $H\chi^3$. Of these, $H^3\chi$ can only be
induced by quark loop diagrams and here too the top loop would dominate.
The expected $H^3\chi$-coupling is ${g_y(3/4)^3\over 8\pi^2}\simeq
10^{-3}$. Likewise the $H^2\chi^2$ induced by W-loop is $\simeq
{(g^2)^2\over 8\pi^2} ln m_H^2/M_W^2\simeq 1/1600$ while that induced by
top loop is ${(g_y)^2(3/4)^2\over 8\pi^2}\simeq 1/3600$. Thus we
conclude that the induced mixing between Higgs and the chiral multiplet
can be safely neglected.
\section{Massless Scenario}
In this section we consider the scenario where $M_{\chi}$ is
small. Physically, this circumstance could either arise when the
$SU(2)_L\otimes SU(2)_R$ is spontaneously broken giving rise to Goldstone
bosons with small masses due to some explicit breaking by quark mass
terms, or when the $SU(2)_L\otimes SU(2)_R$ symmetry is manifest with a
small chiral invariant mass term for the scalars. In the following we analyse
how this scenario is constrained by data on R-parameter, $g-2$ for muons and high precision
Z-width data.
\subsection {	The R parameter}
	
	The R parameter measures  the ratio of 
		  $ \sigma ( e^+e^- \rightarrow hadrons )$ to $ \sigma (e^+e^-
		  \rightarrow \mu^+\mu^-)$. 
In QCD , at high energies, the former is approximated by 
	    $ \sigma( e^+e^- \rightarrow \Sigma q\bar q )$.
When the
energies are well below the Z mass we can neglect the contribution
of the Z mediated processes to R. To leading order , then, the R parameter
measures the total number of operational flavours
(of course, multiplied by the number of colors of the quarks ), 
modulated by their charge squared .

As we move to
higher generations the  R parameter changes rapidly at quark mass
thresholds. At thresholds we also encounter many resonances which also
give rapid changes in the R parameter. However, away from thresholds R
can be quite stable. Therefore in the region above $b\bar b$
threshold we expect R to be stable apart from QCD corrections. But as we
approach  $\sqrt s  = M_z$, a new class of diagrams become operative and 
subsequently the R
parameter has a steady rise to the Z peak.

 There is thus an energy region  , 20- 40Gev , where the effect
 of the Z is yet very small , where R is relatively stable. To leading order
 , that is , in the absence of QCD corrections, here
				$R_0 =  11/3$.

The R parameter will change for our theory as we have new scalar charged
partons , the $\tilde\pi^+\tilde\pi^-$ that couple to the photon which will
contribute to the hadronic cross section. For the the zero mass partons
considered by us the contribution to the R parameter is exactly
calculable and is given by
		  $R = 1/4$.
This additional contribution should be clearly visible, particularly in the region:
20 - 40 Gev . The contribution of Z - exchanges are negligible in this
region.

The R parameter receives QCD corrections and  
the QCD corrected R parameter in this region is:
               $R(s) = R_0 ( 1 +  \alpha_{s} /\pi +...       )$
The measured value of the R parameter in this region has a world average of
4.02. The difference between this number , 4.02 , and  $R_0 = 11/3$ is supposed
to come from the QCD corrections and yield a value for $\alpha_{s}$ 
at  this energy scale.

However if we add the extra contribution of our new partons,
		$R_0  = 11/3 + 1/4  = 3.91$,
leaving only a deficit of .11 to be accounted for by the QCD corrections. The
corresponding value of the QCD coupling will then be too low to be admissable.
However  the systematic errors at Amy, Topaz etc are not so small , of the order of
5\% . Also, different groups have reported  R in this region to be as high as 
4.2 or even as low as 3.8 . This circumstance means that our theory cannot be ruled out
as the effect we are considering is 
				 $\Delta R/ R  =  6$ \%.
It is worth pointing out that experimentally low multiplicity ($\simeq
5$) jets are not counted as these are very unlikely in QCD. On the
other hand, for the color neutral pionic partons of our theory, we
expect to have only 
low multiplicity jets most of which would have been
excluded by experimental cuts.\footnote{ This exemplifies the point made
in the introduction whereby biases based on a particular theoretical
framework
( QCD, in this case ) can surreptiously influence the way the data is
analysed.} Of course, two prong events would have
been counted as $\mu^+\mu^-$-pairs and could have shown up as
anomalies which could be distinguished by their different angular 
dependence 
as compared to  leptons.

Given these facts the R parameter is not at present a definitve test
for this theory versus QCD  though reduced systematic errors could 
put things on the borderline.

\subsection{g-2 for the muon}

 The contribution to g-2 , the muon magnetic moment, can be potentially disturbed
 by the presence of additional charged particles. 
 Our theory, in the light mass scenario considered in this section, has particles with zero or light
 masses coupling to photons thereby contributing additionally to the photon
propagator. Such
 a contribution to the propagator can be related to the contribution to the
 R parameter we have just considered . This is precisely how g-2
 is calculated in the literature \cite{ynd}. In this work
 the contribution to g-2 for the muon , $a_v$ , is given by
 \beeq
                   a_v =\int_{4m^2} dt K(t)R(t)
 \eneq
where $R(t)$ is the R-ratio and $K(t)$ is a known function.

  However , the manner in which zero mass scalar partons enter the low energy description 
  of R(t) ( where $\sqrt t$ is the centre of mass energy ) is subtle. At low energy there is  
  mixing 
  between our zero mass partonic pions and the pseudoscalar channel . 
  This will generate a higher mass state with quantum numbers of the
  pion in addition to the the usual 'goldstone pion ' associated with the spontaneous breaking of
  chiral symmetry. 
  Since we cannot calculate the mass of this non-perturbatively
  generated extra state 
  we can only use experiment to glean its mass. The particle data book lists the first 
  additional state with pion quantum numbers at 1.3 Gev. At low energies then
  we must use this state  to calculate the extra contribution to
  the photon self energy or the R parameter as the usual pion's
  contribution is already accounted for in the standard treatment of
  hadronic contributions to g-2. The threshold for the contribution of this state
 then starts at $\simeq $2.6 Gev. This puts us in the perturbative QCD regime. 

  Before computing the extra contributions a few remarks are in order: 
1) The low energy regime, 0.8-2.0 Gev, has to be gleaned from experiment as perturbative 
QCD can not be used . The
contribution to $a_v$ from this region  as listed in  Table 2  of \cite{ynd} is
	    $(1404 \pm 100 ) . 10^ {-11}$
As this is evaluated from experimental data, one cannot differentiate the 
contributions from QCD and our theory .
2) Perturbative QCD is used for $t >2 Gev^2$ ( Table 1 of
\cite{ynd}) except
for the threshold regions  which are populated by numerous 
resonances,
where again one has to only rely on experimental data, 
and can not differentiate between our theory and QCD.
These regions are:
3.3- 3.6 Gev, 3.6 - 4.9 Gev and  9 - 14 Gev.

  Thus it is only for regions for which estimates  are made via
  perturbative QCD that comparison between this theory and QCD is possible.
In respect 
  of the foregoing discussion this region for us must begin at  $\sqrt t >2.6$ 
  Gev. 
  We briefly sketch how the additional contribution can be evaluated 
for our theory in the 
  region  2.6 - 3.1 Gev:
  i) K(t)  goes as  $1/t^2$. Assuming $R(t) =  R_0$ and using eqn(10)
 we can get the two contributions   for 1.4-2.6 Gev and  2.6-3.1 Gev
for QCD.
The partial contribution for the region , 2.6 -3.1 Gev is found to be 
1/10 of the total.
ii) The extra contribution to R assuming zero mass pionic partons is  
$\delta R$ = 1/4 whereas the
QCD contribution is  $R_0  = 2$.
Thus the fractional extra contribution is  1/8.
iii) This is further down when we take into account the mass of the excited 
state (1.3 Gev). A rough 
estimate is provided by mutiplying by the phase space factor 
$( 1  -   4 M ^ 2/t ) $  =  0.17
taken at the average value 2.85 Gev for $\sqrt t$.  
The total extra contribution  for the region 2.6-3.1 Gev works out to
 $1.2 \cdot 10^{-11}$. Below we display the contributions and errors 
 for various regions:
\begin{tabular}{lllll}

$~~~\sqrt t~~~$ &  ~~~    QCD~~~   &~~~         Chiral~~~    &~~~         Theo.~~~   &      ~~~    Sys.~~~\\
in Gev	       &	      & multiplet                         &           error    &     error($ 5\%$ )\\

 2.6 - 3.1   & ~~~       56              &         1.2                   &       $\pm 1$        &   $ \pm$    2.8\\

 4.9 -  9     &  ~~~      67.5            &         4                     &       $ \pm 1$        &   $\pm$    3.5\\

   $>$  14       & ~~~       13              &         0.9                   &       $\pm 2$        &   $  \pm$   0.65\\

\end{tabular}

   It should be noted that:
   i) The theoretical error in \cite{ynd} is arbitrarily estimated as 
   half the $\alpha_{s}^2$
   correction to R.
   ii) we have taken the systematic error to be  5\% of R (see Sec 4.1)

 The extra contribution of our theory  falls within the sum of 
   the theoretical and the systematic errors .
 The error in the low energy region , 
   0.8-2 Gev is roughly $100\cdot  10^{-11}$. By comparison, all our extra 
contributions are negligible.

   We are therefore led to the conclusion that  g-2 for the muon despite being a very high
   precision measurement of the charged-particle content of theories 
cannot differentiate between QCD and our 
   theory  -  a rather non trivial result.

\subsection{   Z width}

The Z-width data on the other hand is known with great accuracy.
The minimal coupling of the chiral multiplet to $Z_{\mu}$ 
is ( see eqn (6,7)):
\beeq
{\cal L}_{lin}^{neut}= e(A_{\mu}-{\gamma\over 2} Z_{\mu})
(\vec{\tilde\pi}\times
\partial_\mu\vec{\tilde\pi})_3
-{e\over 2cs}Z_\mu(\tilde\pi_0\partial_\mu\tilde\sigma-\tilde\sigma
\partial_\mu\tilde\pi_0)
\eneq
where $\gamma = (1-2s^2)/cs)$ with $s$ being $sin \theta_W$ and $c^2 = 1-s^2$. 
The contribution to the hadronic width of Z-boson due to the extra scalars can be calculated easily:
\beeq
{\Delta\Gamma^Z\over \Gamma^Z_{had}} = {9((1-2sin^2 \theta)^2+1)\over N_c(90-168sin^2\theta + 176 sin^4\theta)}
\eneq
At $sin^2 \theta \simeq .25 $, this works out to roughly 4.5\% of the total width $\Gamma^Z$. The high precision LEP data only allows total uncertainty of about
.3\%.\par 
This immediately rules out the extended version where the
chiral symmetry in the extended sector is spontaneously broken or where the
chiral symmetry in the extended sector is manifest with low mass for the
multiplet, or where chiral symmetry is fully broken but with light masses
for the chiral scalars.

   In conclusion, by explicit construction of an alternative theory to QCD for the strong interactions we
   have found that all precision tests for QCD except forthe Z width cannot select between the 
   two theories. Only by extending the theories to the FULL electrweak standard model do we find
   an unambiguous support in favour of QCD from the Z width.
   This underscores the fact that most tests and vindications of QCD that are to be found in 
   archival refrences in the literature are just not adequate and that 
   only the Z width is precise enough to be the final arbiter\cite{ztest}.

  A different version of this theory  where the chiral multiplet mass
  of
  more than one half the  Z mass will be exempt from this problem 
  This is considered in the next section.

\section{Massive Scenario}
The main conclusion from our analysis of the massless scenario is that
$M_{\chi}< M_Z/2$ is not viable. In this section we consider the opposite
scenario that $M_{\chi} > M_Z/2$. Before looking at the signatures for such
a scenario, we first investigate the constraints on the model by Flavour
Changing Neutral Currents(FCNC) as well by the oblique parameters S, T \& U.

\subsection{Flavour Changing Neutral Currents}
In the standard model $\Delta F = 2$ processes like the ones mediated
by

\begin{figure}[htb]
\begin{center}
\mbox{\epsfig{file=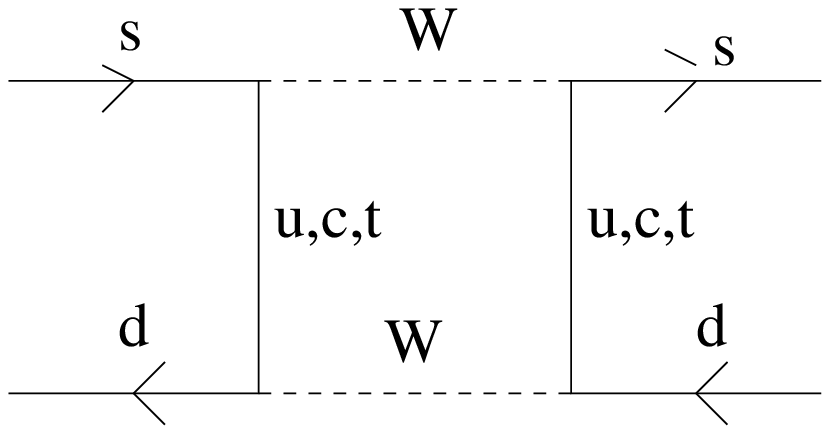,width=3truecm,height=2truecm,angle=-90 }}
\caption{ Typical box diagram contribution to $\Delta F = 2$ processes}
\label{Fig 6.}
\end{center}
\end{figure}
are severely suppressed compared to their generic values. While these
processes could be generically of order $ G_F\alpha$, they actually turn
out to be of order $G_F^2$. This remarkable suppression, borne out by
data, is due to the unitarity of the so-called CKM matrix for three
generations. In our model too FCNC processes must be likewise
suppressed; otherwise they are immediately ruled out. A natural question
to ask is whether there is an analogue of the CKM matrix for our model
too.
As in the standard model  we introduce mass-eigenstate quark-fields through
\beeqar
\psi_R & = & T^{\dag} \psi_R^{\prime}\nonumber\\
\psi_L & = & S^{\dag} \psi_L^{\prime}
\eneqar

Now it is important to see whether FCNC are introduced in our model. We
must have
\beeq
{\cal L}_{yukawa} = F_{AB}\bar \Psi_A^{\prime} (\sigma = i\gamma_5 \vec
\tau\cdot \vec\pi)\Psi_B^{\prime}
\eneq
in order that the interaction preserve $SU(2)_L$- invariance. Here
$\Psi$
is the Dirac field. On splitting the Dirac field into its left and
right-handed components and rotating them into the mass-eigenstate
basis one obtains
\beeqar
& &(S_{(p)}^{\dag} F T_{(p)})_{AB} \bar p_L^A (\sigma + i\pi^0)p_R^B+
(S_{(n)}^{\dag} F T_{(n)})_{AB} \bar n_L^A (\sigma - i\pi^0)n_R^B\nonumber\\
&+ &i\pi^-(S_{(n)}^{\dag} F T_{(p)})_{AB} \bar n_L^A  p_R^B+
i\pi^+(S_{(p)}^{\dag} F T_{(n)})_{AB} \bar p_L^A  n_R^B
\eneqar
Now there are potential FCNC terms even at tree level because of the
first two terms in the above expression.
\begin{figure}[htb]
\begin{center}
\mbox{\epsfig{file=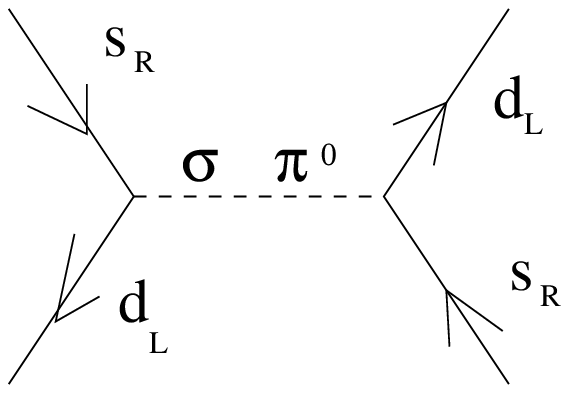,width=3truecm,height=2truecm,angle=-90}}
\caption{ Tree Level contribution to $\Delta F = 2$ processes}
\label{Fig 7.}
\end{center}
\end{figure}
However, the contributions to 
tree level from FCNC due to $\sigma$ and $\pi^0$ exactly cancel, and
 {\bf
there are no tree level FCNC in this model}. This is true both for 
$\Delta F = 2 $ and $ \Delta F = 1 $ processes. This is unlike the
situation
in generic two Higgs doublet models, where often the couplings have to
be fine-tuned to keep tree-level FCNC under control. This is because in such models, the
analogs of $\sigma$ and $\pi^0$ are not mass-eigenstates; instead, the
mass-eigenstates are linear cominations of $(H_0, \sigma, \pi^0)$ where
$H_0$ is the higgs of the minimal standard model. This is again due to
the fact that the most general Higgs potential that is $SU(2)\times
U(1)$-invariant, allows such a mixing(of course the analogs of
$(\pi^{\pm})$ are mass-eigenstates even in such models). On the other
hand in our model the Higgs-chiral multiplet coupling term has to be
$SU(2)_L\times SU(2)_R$-invariant, which only allows
\beeq
|\Phi|^2\cdot (\sigma^2 + \vec \pi^2)
\eneq
and this keeps $(\sigma, \pi^0)$ as mass eigenstates even after SSB,
except  for the small mixing effects described in sec.3.3.

This cancellation of FCNC due to the tree-level $\Delta F = 1$ vertices
persists even for the one-loop contributions of the type shown below
\begin{figure}[htb]
\begin{center}
\mbox{\epsfig{file=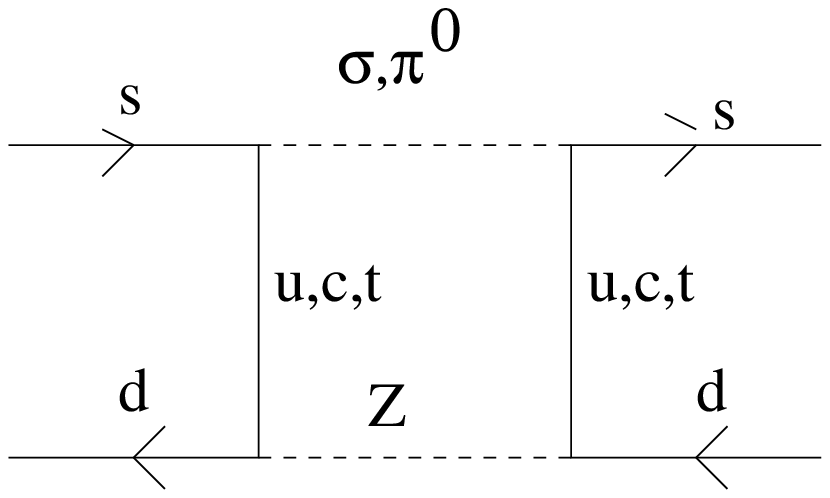,width=3truecm,height=2truecm,angle=-90}}
\caption{ Neutral exchange contribution to $\Delta F = 2$ processes}
\label{Fig 8.}
\end{center}
\end{figure}
The FCNC generated by the last two terms of eqn() are more complicated
and involve deeper analysis. The generic contributions to 
$\bar d s \rightarrow \bar s d$ 
involve box diagrams of two types:a) $ W\pi$
exchange and b)$\pi \pi$ exchanges shown below
\begin{figure}[htb]
\begin{center}
\mbox{\epsfig{file=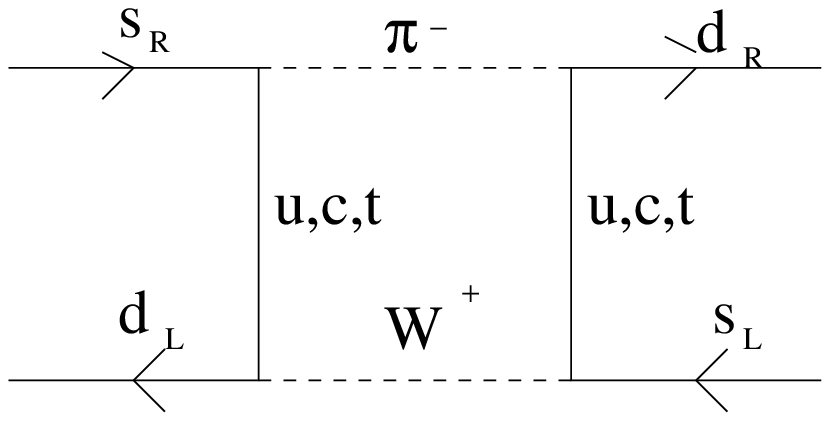,width=3truecm,height=2truecm,angle=-90}}
\label{Fig 9.}
\end{center}
\end{figure}
\begin{figure}[htb]
\begin{center}
\mbox{\epsfig{file=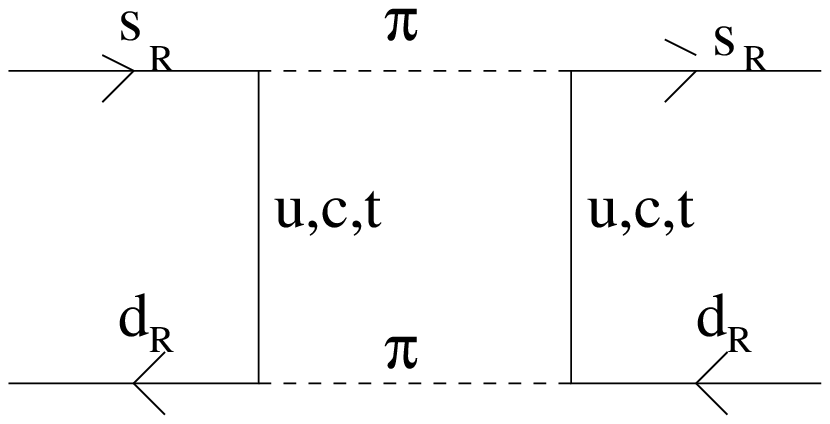,width=3truecm,height=2truecm,angle=-90}}
\label{Fig 10.}
\end{center}
\end{figure}
In addition, unlike the standard model case, there are several $\Delta F
= 2$ operators to deal with. We shall illustrate this for the particular
case of 
$\bar d s \rightarrow \bar s d$ by adopting the notation that 
$\bar d_{\lambda_1} s_{\lambda_2} \rightarrow \bar s_{\lambda_3} d_{\lambda_4}$ shall be denoted by 
$(\lambda_1, \lambda_2, \lambda_3, \lambda_4)$. Thus the case of fig.1 is $(L,
R, L, R)$.In principle
16 possibilities are possible.But for the $w\pi$ exchange, at the
W-vertex only L's are allowed. This way the number of possibilities is
cut down considerably. Also, diagrams where the internal i-quark propagator
connects quark fields of different helicities are suppressed by
$m_i^2/M_W^2$. Thus the independent classes to be counted are: for the
$W\pi$-exchange, (L, R, L, R) and for the $\pi\pi$-exchanges, $(L, L, \lambda,
\lambda)$ and $(R,R,\lambda,\lambda)$.

The analysis is facilitated on noting that there are four coupling
matrices in eqn() denoted by
\beeqar
V^p_{neut} 
& = & S_{(p)}^{\dag} F T_{(p)}\nonumber\\
V^n_{neut} 
& = & S_{(n)}^{\dag} F T_{(n)}\nonumber\\
V^p_{char} 
& = & S_{(p)}^{\dag} F T_{(n)}\nonumber\\
V^n_{char} 
& = & S_{(n)}^{\dag} F T_{(p)}
\eneqar
It should be noted that 
$V^p_{char}$ is associated with outgoing $p_L$ and incoming $n_R$ (the
indices are also oredered this way), while
$V^n_{char}$ is associated with outgoing $n_L$ and incoming $p_R$. It is
further useful to note
\beeqar
V^p_{char} & = & V_{CKM} \cdot V^n_{neut}\nonumber\\
V^n_{char} & = & V_{CKM}^{\dag} \cdot V^p_{neut}
\eneqar
The contributions to various $\Delta S = 2$ processes can be summarised
thus
\beeqar
{(L, R, \lambda_1, \lambda_2)}_{W\pi} & \simeq & (V_{CKM}^{\dag}V^p_{char})_{12} =
				       (V^n_{neut})_{12}\nonumber\\
{(R, R, \lambda_1, \lambda_2)}_{\pi\pi} & \simeq & ({V^p_{char}}^{\dag}V^p_{char})_{12} =
				       ({V^n_{neut}}^{\dag}V^n_{neut})_{12}\nonumber\\
{(L, L, \lambda_1, \lambda_2)}_{\pi\pi} & \simeq & ({V^n_{char}}^{\dag}V^n_{char})_{12} =
				       ({V^p_{neut}}^{\dag}V^p_{neut})_{12}
\eneqar
If we had considered FCNC for $\bar u_R c_R \rightarrow \bar c_R
u_R$, the
first of these eqns would habe been modified by the
replacement$V^n_{neut} \rightarrow V^p_{neut}$.

Thus we see that natural suppression of all FCNC operators would
require that the matrices
$V^p_{neut},  
V^n_{neut}$ are both {\bf diagonal}. Now we use two theorems(which are
quite easily proved) to show that this can happen {\bf only} when the
matrix F is proportional to the Identity-matrix.

{\bf Theorem 1}\\
If a hermitean matrix M is  diagonalised by a bi-unitary
transformation
\beeq
VMU^{\dag} = M_d
\eneq
then 
\beeq
V U^{\dag} = W_d
\eneq
where$W_d$ is some diagonal unitary matrix.

{\bf Theorem 2}\\
If a Hermitean matrix M is brought to diagonal form by two sets of
bi-unitary transformations
\beeqar
V_1 M U_1^{\dag} & = & W_d^1 H_d\nonumber\\
V_2 M U_2^{\dag} & = & W_d^2 H_d
\eneqar
and if
\beeq
V_1 V_2^{\dag} \ne W^3_d
\eneq
or
\beeq
U_1 U_2^{\dag} \ne W^4_d
\eneq
where $W^{(3, 4)}_d$ are some diagonal unitary matrices,
then
\beeq
M = m I
\eneq
Since in our case F is a hermitean matrix, theorem-1 immediately
implies 
\beeqar
S_{(p)} T_{(p)}^{\dag} = W^{(1)}_d\nonumber\\ 
S_{(n)} T_{(n)}^{\dag} = W^{(2)}_d
\eneqar
As stressed earlier, natural suppression of {\bf all} FCNC in this
model would require that $V^p_{neut}, V^n_{neut}$ are both diagonal. From
the theorem proved above it is also clear that these have to be
diagonal unitary matrices.

In our case $S_p^{\dag}S_n$ is the CKM-matrix which is not diagonal;
hence theorem-2 implies {\bf F must be a multiple of unit matrix}.

This in turn implies that 
\beeqar
V^p_{neut} & = & W_d^1\nonumber\\
V^n_{neut} & = & W_d^2\nonumber\\
V^p_{char} & = & V_{CKM} \cdot W_d^2\nonumber\\
V^n_{char} & = & V_{CKM}^{\dag} \cdot W_d^1
\eneqar
where $W_d^{1, 2}$ are some diagonal unitary matrices.

This is the unique solution to natural FCNC suppression in this
model. It is instructive to pause and reflect whether this manner of
FCNC suppression amounts to fine-tuning in this model. A particular
solution is not a fine-tuned solution if it enhances the symmetries of
the system. In the absence of $SU(2)\times U(1)$ couplings our
solution indeed enhances the symmetry from $SU(2)_L\times SU(2)_R$
to 
$SU(3)_{hor}\times SU(2)_L\times SU(2)_R$. 
Thus if the standard
model is not truly fundamental but only an effective description, then
there could be an intermediate phase(at scales above the weak scale)
where nature may have preferred 
$SU(3)_{hor}\times SU(2)_L\times SU(2)_R$. If that is so, the
chiral-multiplet may be even more fundamental than what we may have
been thinking.

Lastly, we see that our solution to the suppression of FCNC can not
generically yield CP-conservation. The minimal CP-nonconservation can be
achieved by taking $W_d^{(1, 2)}$ to be unit matrices.

\subsection{The electroweak precision parameters S, T and U 
for the extended theory }
The so called oblique parameters 
\footnote{The results of this section were obtained in
collaboration with Dr. Rahul Sinha. We are also indebted to him for many
useful discussions.}
as precision tests for electroweak
theories are defined by \cite {pesk,prob,hagi}
\beeqar
\alpha \tilde T M_W^2 &=& \tilde\pi_{WW}(0) - c ^2  
\tilde\pi_{ZZ}(0) \nonumber \\
\alpha \tilde S M_Z^2 &=& 4 cs [cs (-\tilde\pi_{ZZ}(0) - 
\tilde\pi_{\gamma\gamma} (M_Z^2)) + (s^2 - c^2)\tilde\pi_{\gamma Z}(M_Z^2)]
\nonumber \\
\alpha \tilde U M_W^2 &=& 4 s^2 [c^4 \tilde\pi_{ZZ}(0) - \tilde\pi_{WW}(0)
\nonumber\\
&-&c^2s( s\tilde\pi_{\gamma\gamma} (M_Z^2) + 2 c \tilde\pi_{\gamma Z}(M_Z^2))]
\eneqar

In these equations $\tilde\pi_{VV}$ refers to the propagator function for
the vector boson V after mass and wavefunction renormalisations. For
the mixed propagator function $\tilde\pi_{\gamma Z}$, the mass and
wavefunction renormalisations are understood to be carried out on the
photon pole; c and s stand for $cos \theta_W$ and $sin \theta_W$
respectively.

It turns out that all propagator functions $\tilde\pi_{AB}$ before mass
and wavefunction renormalisations are proportional to each other, so in
this section we shall first consider the details of that common
function.

Denoting the generic one-particle irreducible 2-point function by ${\cal
M}_{\mu\nu}$, one eventually finds
\beeq
{\cal M}_{\mu\nu} = -iAg_{\mu\nu}+iBP_{\mu}P_{\nu}
\eneq
with
\beeq
A = B P^2 = \Pi(P^2)
\eneq
where
\beeqar
\Pi(P^2) &=& -{1\over 48\pi^2}(\gamma_E+ln M_\pi^2)P^2\nonumber\\ 
&+& {P^2\over 16\pi^2}({2\over 9}-
{8\over 3}a^2+{16\over 3}a^3~arctg ({1\over 2a}))
\eneqar
Thus the full two point function has the expected
gauge-invariant structure
\beeq
{\cal M}_{\mu\nu}= -i(g_{\mu\nu}-{P_\mu P_\nu\over P^2})\Pi(P^2)
\eneq
Recall that the various $\tilde\pi(P^2)$ are mass and wavefunction
renormalised. Because $\Pi(0)=0$, mass renormalisations for
$\tilde\pi_{\gamma\gamma}, \tilde\pi_{\gamma Z}$ are automatically taken
care of. To discuss wavefunction renormalisation, we compute
$\Pi(P^2)^{\prime}$ where $\prime$ denotes differentiation wrt $P^2$. The result is
\beeq
\Pi(P^2)^{\prime} = -{1\over 48\pi^2}(\gamma_E+ln M_\pi^2)+{1\over
18\pi^2}+{f_2(a)\over 16\pi^2}
\eneq
where
\beeq
f_2(a) = {8a\over 3}[-(a^2+3/4)arctg({1\over 2a}) +{a\over 2}]
\eneq
introducing
\beeq
f_1(a) = 2/9-8/3a^2+16/3 a^3~arctg(1/2a)
\eneq
we can recast $\Pi(P^2)$ as
\beeq
\Pi(P^2) = {P^2\over 16\pi^2}[-1/3(\gamma_E+ln M_\pi^2)+f_1(a)]
\eneq
It should be noted that $f_1(a)_{a\rightarrow\infty}\rightarrow 0$. The
fully renormalised quantities renormalised at $P^2 = M^2$ denoted by
$\tilde\pi(P^2)^{(M)}$ are given by
\beeq
\tilde\pi(P^2)^{(M)} = \Pi(P^2)-\Pi(M^2)-(P^2-M^2)\Pi(M^2)^{\prime}
\eneq
The case $M=0$ will be explicitly worked out as the
limit $a\rightarrow\infty$ could be problematic for a numerical
evaluation. The result is
\beeq
\tilde\pi(P^2)^{(0)} = {P^2\over 16\pi^2}[f_1(a)-2/3]
\eneq
The result for the generic case is
\beeqar
\tilde\pi(P^2)^{(M)} &=& {P^2\over 16\pi^2}f_1(a)-{M^2\over 16\pi^2}f_1(a_M)\nonumber\\
&-&{(P^2-M^2)\over 16\pi^2}(f_2(a_M)+8/9)
\eneqar
where $a_M^2= M_\pi^2/M^2-1/4$.
Now we only have to identify the coupling constants that multiply the
various functions:
\beeqar
\tilde\pi_{WW} &=& {e^2\over 2s^2}\tilde\pi^{(M_W)}\\
\tilde\pi_{ZZ} &=& e^2({\gamma^2\over 4}+{1\over 4s^2 c^2})\tilde\pi^{(M_Z)}\\
\tilde\pi_{Z\gamma} &=& -{e^2\gamma\over 2}\tilde\pi^{(0)}\\
\tilde\pi_{\gamma\gamma} &=& e^2\tilde\pi^{(0)}
\eneqar
here $e^2 = 4\pi\alpha$ and $\gamma = (1-2s^2)/sc
$  with $s^2 =
0.23$.

It should be noted that the structure of eqn (52) is such that
$\tilde\pi^{(0)}$ is evaluated at $M_Z^2$ and $\tilde\pi^{M_Z, M_W}$
is evaluated at $P^2=0$. Hence we can use
\beeqar
\tilde\pi^{(0)}(M_Z^2) & = & {M_Z^2\over 16\pi^2}[f_1(a_Z)-2/3]\nonumber\\
\tilde\pi^{(M)}(0) & = & {M^2\over 16\pi^2}[f_2(a_M)-f_1(a_M)+8/9]
\eneqar
We finally give here the corrections to
these parameters:
\vspace{0.1in}
\begin{center}
\begin{tabular}{llll}
\hline
\hline\\
Scalar Mass(GeV) & $~~~  \delta\tilde T$& $~~~    \delta\tilde S$ & 
$~~~    \delta\tilde U$ \\ 
\hline
\hline\\
 ~~~~~~~~     50 &-.010  & -.027  & .005   \\ 
 ~~~~~~~~     55 &-.005  & -.018  & .002   \\ 
 ~~~~~~~~     60 &-.003  & -.013  & .001   \\ 
\hline
\hline\\
\end{tabular}
\end{center}
these corrections are much smaller than the uncertainties in even the
most precise LEP measurements \cite{hagi}.

\subsection {Other Precision Tests for the Model}:
The coupling of the chiral multiplet to the electroweak bosons
given by eqns(6-7) leads to additional contributions to various 
processes of the electroweak theory. One of the sensitive tests for QCD
is the value of the R-parameter. As the cross-section for the pair
production of the scalars at LEP energies is of the order of a 1pb, the
R-parameter is not very sensitive to the presence of the scalars. The
other important precision test is the g-2 for muons.
There are two types of additional contributions
to g-2 that arise. One is due to the enhanced ultraviolet degrees of
freedom
and the other due to  additional hadronic
interactions . The contributions due to the former 
arise out of the
modification of the photon propagation function 
The analytic result is 
$\Delta g = {\alpha\over 2\pi}{\alpha\over 180\pi}{m_{\mu}^2\over m_{\tilde\pi}^2}$
For $m_{\tilde\pi} = 45 GeV$ this amounts to $\Delta g = 4\cdot 10^{-14}$ and hence
insignificant. The shift due to the modification of hadronic interactions is
much harder to estimate precisely.  
One should expect very little difference between QCD and the 
extended theories here because of the expected decoupling of massive
particles. 
One expects the additional interactions
to produce changes in g-2 at the level of ${g_y^2 \over 4\pi} {1\over 2\pi}$
times the dominant hadronic contributions. This amounts to less than 2 parts in
 1000 of the dominant hadronic contributions and is hence much less than the
known theoretical uncertainties in g-2. 

\section {Four Jet Events }
As mentioned in the introduction, though the ALEPH collaboration has now
retracted its earlier claims of having seen excess four jet events, many
features reported by them earlier follow naturally and in a virtually
parameter-free manner from our model. We find it worthwhile to present them
here as generic features of four jet events for our model.

In our model one expects excess four jet events identified with 
the decay products of the
scalars which have no couplings to the leptons. Hence one of the
characterstic features of such four jet events in our model are:
a) no leptons with high transverse momentum wrt the
jets are expected.
b) no events of the type $\tau^+\nu_{\tau}\bar cs$ and 
$\tau^+\nu_{\tau}\tau^-\bar\nu_{\tau}$ are expected.

The following table gives in pb the cross-sections
for $e^+e^-\rightarrow \tilde\sigma\tilde\pi^0 (\tilde\sigma_{neut})$ and to 
$\tilde\pi^+\tilde\pi^-(\tilde\sigma_{char})$ as a function of $\sqrt s$ and scalar mass. Taking 
into account the FCNC constraint that the scalars couple nearly equally
to all flavours, the branching ratios for having at least 2 b($\bar b$)
jets ($R_{2b}$) and four b($\bar b$) jets ($R_{4b}$) are also given. 

\begin{center}
\begin{tabular}{llllll}
\hline
\hline\\
$\sqrt s$(GeV) & $~~~  M_{\tilde\pi}$& $~~~ \tilde\sigma_{neut}$ & $~~~   \tilde\sigma_{char}$
&~~~~$R_{2b}$&$~~~~R_{4b}$ \\ 
\hline
\hline\\
 ~~~~~130 &~~~50  &~~~ .63  &~~~ .54&~~~1/5&~~~1/46   \\ 
 ~~~~~130 &~~~55  &~~~ .37  &~~~ .31 &~~~1/5&~~~1/46  \\ 
 ~~~~~130 &~~~60  &~~~ .14  &~~~ .12 &~~~1/5&~~~1/46  \\ 
 ~~~~~161 &~~~50  &~~~ .43  &~~~ .57 &~~~2/13&~~~1/58  \\ 
 ~~~~~161 &~~~55  &~~~ .35  &~~~ .46 &~~~2/13&~~~1/58  \\ 
 ~~~~~161 &~~~60  &~~~ .27  &~~~ .35 &~~~2/13&~~~1/58  \\ 
 ~~~~~172 &~~~50  &~~~ .38  &~~~ .55 &~~~1/7&~~~1/61  \\ 
 ~~~~~172 &~~~55  &~~~ .32  &~~~ .46  &~~~1/7&~~~1/61 \\ 
 ~~~~~172 &~~~60  &~~~ .26  &~~~ .37 &~~~1/7&~~~1/61  \\ 
\hline
\hline\\
\end{tabular}
\end{center}
In our model the widths of the scalars are :
$\Gamma^{\tilde\sigma, \tilde\pi^0}_{\bar q q} =
{g_y^2\over 4\pi}~{3M_{\tilde\sigma}\over
2}$ and
$\Gamma^{\tilde\pi^+\tilde\pi^-}_{\bar q q} =
{g_y^2\over 4\pi}~3M_{\tilde\sigma}$.
At $\rho = 1/36$ , the former amounts to about
0.375 Gev/flavour, and with a top mass
of 170 Gev, only five flavours
contribute to the decay of
neutrals, leading to a width of
about 1.88 Gev.
Likewise, the latter works out to 0.75 Gev/flavour, and 
the width of charged scalars is
about 1.5 Gev. These should be the expected widths for the di-jet mass-sum
distribution.
We present a comparison of
the expected properties of four jet events in our model with what one
would expect from some other models in the following Table.

\begin{table}
\caption{COMPARSION}
\vspace{0.1in}

\begin{tabular}{llll}
      & SUPERSYMMETRY& TWO HIGGS    & OUR MODEL \\ 
\hline\\
Additional      &Charginos, Squarks& $ HA,  H^+H^-$ & $\tilde\sigma, \vec{\tilde\pi}$   \\
Particles      & Neutralino...    &                &            \\
\hline
Additional Parameters & many & 2 Yukawa, 5 Self  & 1 Yukawa, 2 Self  \\
\hline
$\sigma_{tot}$(pb)(Lower $\sqrt s$) & 1 to 7  & $\sim 1 $ & $\sim 1.17 $  \\ 
(Higher $\sqrt s)$&similar&similar &$\sim 1 $  \\ 
\hline
Width Of   & Large & Depends on & 1 to 2 Gev  \\
Mass Dist  &          & Yukawa coupling &     \\
\hline
Final State& Expected & Expected& No leptons  \\
 Leptons  &           &                 &     \\
\hline
$R_{2b}$ &  Expected &   Predominant & 1/5-1/7 with 100\% eff \\
\hline
$R_{4b}$ &  Expected &   Predominant & 1/45-1/63  \\
\hline
FCNC     & Fine tuning & Fine Tuning & Naturally fulfilled   \\
\end{tabular}
\end{table}

\section{ Other Experimental Signatures}
In this section we shall discuss some other experimental 
signatures for the extended theories that appear feasible at the moment.
\subsection{Real Photon Processes}
In electron-proton(ep) scattering experiments, by restricting to suitable 
regions of scattered electron angles and energies, one can get almost real
photons. We now describe the new events predicted by the extended model in those
regions.\par
The dominant process in ep-scattering is essentially electron-quark scattering
(eq) which manifests itself as a 2 jet event with activity throughout the
rapidity region. The advantages of  working with nearly real photons is that
this dominant process is kinematically forbidden. Indeed, the leading order 
process now is $\gamma q -> qg$ which experimentally manifests itself as 
2+1(the gluon jet, the scattered quark jet and the beam jet)
jet topology with no rapidity gaps. In the extended theory we now have the
additional process $\gamma q -> q\pi(\sigma)$. These events are 
characterised by 2+2 jet topologies where the last two jets 
arise out of the decay of $\pi(\sigma)$, and there will be significant 
rapidity gaps in the $(q, \pi), (\pi, 
beam)$ regions. Now the relative fraction of these events to the dominant QCD
process is $\sim N_g^` \rho = 1/12$ apart from some phase space suppression. 
This is a sizeable effect. 
\subsection{ep scattering}         
In ep scattering the dominant QCD process is the 2 jet event with no rapidity
gap produced by the subprocess $eq -> eq$. Of the 2 jets, the quark jet is 
produced back to back with the electron in the eq centre of mass frame. In the
extended model, the dominant process is $eq -> eq\pi(\sigma)$. The 
distinguishing features of these new events are that only the rapidity region 
between the beam jet and the quark jet is filled. 
Also, the quark jet is no longer back to back with the 
electron in the eq centre of mass frame and can in fact be produced at 
sufficiently small angles. 
\subsection{pp($\bar p$) scattering}         
In addition to the dominant one gluon exchange, now one can exchange the scalar
particles leading to a slightly different angular distribution for jets. The
scalar admixture will roughly be a fraction $N_g^{\prime}\rho$ and could easily
be about 3\%. Another way this could manifest is in a slightly different
estimate for $\alpha_s$ in pp($\bar p)$ reactions compared to $e^+e^-$ or $ep$
reactions.
\section{Acknowledgements}
Much of this work owes its existence to discussions with many people. Important
among them have been our discussions with the experimentalists at KEK notably
K. Abe, Tauchi , Matsui, as well as with the theoreticians  P.M. Zerwas, 
K. Hagiwara, Rahul Sinha, H.S. Sharatchandra, R. Anishetty , R. Basu , G. Rajasekaran,
R. Godbole, Probir Roy, D.P. Roy and Debajyoti Choudhury.
An important part of this work was done while both of us 
were visiting KEK, Japan. Our thanks are to the Theory Group there for its
hospitality.




\begin{thebibliography}{99}
\bibitem[\dag]{email}{\em electronic address :} dass@imsc.ernet.in
\bibitem[\ddag]{email}{\em electronic address :} vsoni@ren.nicnet.in
\bibitem{free} G. t'Hooft, in Marseille Conference on Renormalisation of Yang-Mills Fields, June 1972;
D. Gross and F. Wilczek, Phys. Rev. Lett. {\bf 30} (1973) 1343; H.D. Politzer, Phys. Rev. Lett. {\bf 30} (1973)
1346.
\bibitem{seiler}E. Seiler and A. Patrasciou, "The Problem of Asymptotic Freedom", hep-ph/9609292;
Phys. Rev. {\bf D 57} (1998) 1394-1396.
\bibitem{hooft0} G. t'Hooft, Comm. Math. Phy. {\bf 86} (1982) 449.
\bibitem{gross}S. Coleman and D. J. Gross, Phys. Rev. Lett. {\bf 31}, 851(1973).
\bibitem {spell} G.t'Hooft, in " Under the Spell of the Gauge
Principle", World Scientific 1994, p.244-245.
\bibitem{pre} V. Soni, Mod. Phys. Letts {\bf A 11}, No 4,331(1996)
\bibitem{ztest}N.D. Hari Dass and V Soni, Quantum Chromodynamics and the Z-Width 
,hep-ph/9709464; Mod. Phys. Letts. {\bf A 14}, Nos 8 \& 9(1999), 559.
\bibitem{aleph} N. D. Hari Dass and V. Soni, "Asymptotically Free Alternatives 
to QCD and ALEPH Four Jet events" hep-ph/9709391
\bibitem{schre}B. Schrempp and F. Schrempp, Phys. Lett. {\bf B299}, 3221 (1993).
\bibitem{cheng} "Gauge Theory of Elementary Particles" by T.P. Cheng and L.F. Li,
Clarendon Press, 1984, p.286.
\bibitem{ross}B. Pendelton and G. G. Ross, Phys. Lett. {\bf 98B}, 291(1981).
\bibitem{Hil}C. T. Hill , Phys. Rev. {\bf D 24} (1981) 691
\bibitem{Zim}W. Zimmermann, Comm. Math. Physics {\bf 97} (1985) 211
\bibitem{KSZ}J. Kubo. K. Sibold and W. Zimmermann , Nucl. Phys. {\bf B 259} (1985) 331
\bibitem{tana}V. A. Miransky, M. Tanabashi and K. Yamawaki, Phys. Lett.
{\bf B221}, 177(1989), W. A. Bardeen, C. T. Hill and M. Lindener, 
Phys. Rev. {\bf D41} , 1847(1990).
\bibitem{KTY}K. I. Kondo. M. Tanabashi and K. Yamawaki, 
Prog. Theo. Phys. {\bf 91} (1994) 541
\bibitem{KMZ} J. Kubo, M, Mondragon and G.Zoupanous, Nucl. Phys. {\bf 259} (1994) 291
\bibitem{har}M. Harada ,   Kikkukawa ,  Kugo and  Nakano , Prog. Theo. Phys. 
{\bf 92} (1994) 1161
\bibitem{ynd} J.A. Casas, C. Lopez and F.J. Yndurain  Phys Rev D 32(1985) 736
\bibitem{aleph1}The ALEPH Collaboration, Zeitsschrift f\"ur Physik C71(1996) 179 
\bibitem{rag} F. Ragusa, Status Report before LEP Council, 19 November
1996
\bibitem{dorn} F. Dornier, Status Report before LEP Council, 19 November
1997
\bibitem{haber}"The Higgs Hunter's Guide", J.F. Gunion et al, Addison Wesley,
1990, p 194 
\bibitem{pesk} M.E. Peskin and T. Takeuchi, Phys. Rev. {\bf D46},
381(1972)
\bibitem{prob} A. Kundu and P. Roy, Int. Jour. Mod. Phys. {\bf A12} (1997) 1511-1530.
\bibitem{hagi}K. Hagiwara, Implications of Precision Electroweak Data,
KEK-TH-461, hep-ph/9512425.
\end{thebibliography}
\end{document}